\documentstyle[prc,aps,epsfig,amssymb,preprint]{revtex}
\begin{document}
\preprint{Mz-Th/00-34}
\draft
\title
{\bf MOMENTUM SPACE INTEGRAL EQUATIONS FOR THREE 
CHARGED PARTICLES: DIAGONAL KERNELS} 
\author{\bf A. M. Mukhamedzhanov\thanks{Email: akram@comp.tamu.edu}} 
\address{Cyclotron Institute, Texas A\&M University, College Station, TX 
77843, USA} 
\author{\bf E. O. Alt\thanks{Email: Erwin.Alt@uni-mainz.de}}
\address{Institut f\"ur Physik, Universit\"at Mainz, D-55099 Mainz, 
Germany}
\author{\bf G. V. Avakov\thanks{deceased}}
\address{Institute for Nuclear Physics, Moscow State University, Moscow, 
119899, Russia}

\date{\today}
\maketitle
\begin{abstract}
It has been a long-standing question whether momentum space integral equations of the Faddeev type are applicable to reactions of three charged particles, in particular above the three-body threshold. For, the presence of long-range Coulomb forces has been thought to give rise to such severe singularities in their kernels that the latter may lack the compactness property known to exist in the case of purely short-range interactions. Employing the rigorously equivalent formulation in terms of an effective-two-body theory we have proved in a preceding paper [Phys.\ Rev.\ C {\bf 61}, 064006 (2000)] that, for all 
energies, the nondiagonal kernels occurring in the integral equations 
which determine the transition amplitudes for all binary collision 
processes, possess on and off the energy shell only integrable 
singularities, provided all three particles have charges of the 
same sign, i.e., all Coulomb interactions are repulsive. 
In the present paper we prove that, for particles with charges of equal sign, the diagonal kernels, in contrast, possess one, but only one, nonintegrable singularity. The latter can, however, be isolated explicitly and dealt with in a well-defined manner. Taken together these results imply that modified integral equations can be formulated, with kernels that become compact after a few iterations. This concludes the proof that standard solution methods can be used for the calculation of all binary (i.e., (in-)elastic and rearrangement) amplitudes by means of momentum space integral equations of the effective-two-body type.
\end{abstract}

\newpage

\section{Introduction}
The question of compactness of the Faddeev \cite{fad61} or the equivalent Alt-Grassberger-Sandhas (AGS) \cite{ags67} momentum space integral equations for three charged particles is related to the analytical properties of their kernels. In the preceding paper \cite{maa00}, henceforth called I, we have investigated the analytical behavior of the nondiagonal kernels of the equations for three particles interacting via Coulomb-like pair potentials, rewritten in the form of effective-two-body AGS equations pertaining to all binary (so-called $2 \to 2$) reaction amplitudes \cite{ags67}. Under the assumption that the charges of all three particles are of the same sign, i.e., that all Coulomb potentials are repulsive, their nondiagonal kernels were found to possess only integrable singularities.

In this second part we investigate the singularity structure of the corresponding diagonal kernels, 
again restricting ourselves to purely repulsive Coulomb interactions. It will be shown that the only nonintegrable singularity (i) occurs on the energy shell, and (ii) coincides with the singularity found by Veselova (\cite{ves70}, see also \cite{fad69}) below the breakup threshold and by Alt and Sandhas (\cite{as80} and references therein) for all energies. As is well known, this singularity can be singled out and inverted explicitly \cite{asz78}. All other singularities of the diagonal kernels, including all off-the-energy-shell singularities, turn out to be integrable. Consequently, these equations can be recast in a form such that the kernels of the resultant equations become compact after a few interations, for all energies. 

This completes the investigation of the analytical properties of 
effective-two-body AGS equations for $2 \to 2$ reactions of 
three particles with charges of equal sign above the three-body threshold. 

It is worth mentioning that from the proofs also follows that, as soon as charges with opposite sign are involved, the kernels do, indeed, develop severe singularities which preclude application of standard methods of integral equations theory. This agrees with the findings in the integro-differential approach \cite{fm85}.

The paper is organized as follows. In order to accommodate the reader who is not interested in mathematical details we have collected all relevant definitions and final results in Sec.\ \ref{results}. In particular, in Sec.\ \ref{remarks} we briefly recall the general form of the diagonal kernels of the effective-two body equations as already outlined in \cite{maa00} and discuss some of their pertinent properties. The main finding concerning the leading singularities of the diagonal effective potentials occurring therein is formulated as a Theorem in Sec.\ \ref{ssba}. The resulting singular behavior of the diagonal kernels and its treatment is described in Sec.\ \ref{lskernel}. And in Sec.\ \ref{practical} we sketch the two established solution strategies which eventually lead to the desired physical binary reaction amplitudes of two charged fragments. All proofs of the assertions are deferred to Sec.\ \ref{koff}. Finally, Section \ref{concl} contains concluding remarks. 
 
As this paper is the continuation of Part I, all quantities which are not defined here are given there. And equations of Part I will be referred to as Eq.\ (I.*). 

As usual we choose units such that $\hbar = c = 1$. Moreover, unit vectors are 
denoted by a hat, i.e., ${\hat {\rm {\bf v}}} = {\rm {\bf v}}/v$.

\newpage

\section{Leading singularities of the diagonal kernels
${\cal K}_{\alpha \alpha}$: Resumm$\acute{\rm \bf e}$e} 
\protect\label{results}


\subsection{General remarks}  \protect\label{remarks}

The diagonal kernels occurring in the integral equations which determine simultaneously the transition amplitudes for all binary processes \cite{ags67},
\begin{eqnarray}
{\cal T}_{\beta \alpha}({\rm {\bf q}}_{\beta}', 
{\rm {\bf q}}_{\alpha};z) = {\cal V}_{\beta \alpha}
({\rm {\bf q}}_{\beta}', {\rm {\bf q}}_{\alpha};z) 
+ \sum_{\nu=1}^3\;\int {\frac{d{\rm {\bf q}}''_{\nu}}{({2 \pi})^{3}}} 
\,{\cal K}_{\beta \nu} ({\rm {\bf q}}_{\beta}', 
{\rm {\bf q}}_{\nu}'';z) 
\, {\cal T}_{\nu \alpha} ({\rm {\bf q}}''_{\nu}, \,
{\rm {\bf q}}_{\alpha};z), \label{ags1} 
\end{eqnarray}
are given as 
(see Eq.\ (I.10), with $\beta = \alpha$) 
\begin{eqnarray} 
{\cal K}_{\alpha \alpha} ({\rm {\bf{q}}}_{\alpha}', 
{\rm {\bf{q}}}_{\alpha};z) := {\cal V}_{\alpha \alpha} 
({\rm {\bf{q}}}_{\alpha}', {\rm {\bf{q}}}_{\alpha};z) \;
{\cal G}_{0;\alpha}({\rm {\bf{q}}}_{\alpha};z), \label{dkernel}
\end{eqnarray}
Here, $z=E+i0$ with $E$ being the total energy in the three-body center-of-mass system. 

The diagonal effective potential is defined as 
(Eq.\ (I.11a) with $\beta = \alpha$):  
\begin{eqnarray}
{\cal V}_{\alpha \alpha}({\rm {\bf{q}}}_{\alpha}',
{\rm {\bf{q}}}_{\alpha};z) = \langle {\rm {\bf{q}}}_{\alpha}',
\chi_{\alpha} \mid G^{C}(z) - G_{\alpha}^{C}(z) \mid
\chi_{\alpha},{\rm {\bf{q}}}_{\alpha} \rangle.\label{d1efpot} 
\end{eqnarray}
Note that as a result of assumption (I.7), namely that the short-range interactions are described by separable potentials of rank one (which does not limit the generality of our results as explained in I), it contains only purely Coulombic quantities, {\em viz.} the resolvents $G^{C}(z)$, Eq.\ (I.12), of the three-particle Coulomb Hamiltonian, and $G_{\alpha}^{C}(z)$, Eq.\ (I.13), of the Coulomb channel Hamiltonian. They are related through the resolvent identities 
\begin{mathletters}
\label{resolv}
\begin{eqnarray}
G^{C}(z) &=& G_{\alpha}^{C}(z) + G_{\alpha}^{C}(z) {\bar V}_{\alpha}^{C} G^{C}(z) \label{resolv1} \\ 
&=& G_{\alpha}^{C}(z) + G^{C}(z) {\bar V}_{\alpha}^{C} G_{\alpha}^{C}(z).\label{resolv2} 
\end{eqnarray}
\end{mathletters}
Here, ${\bar V}_{\alpha}^{C}= \sum_{\nu \neq \alpha} V_{\nu}^{C}$ is the Coulomb part of the channel interaction (I.5). Assumption (I.7) implies in addition that in each channel there 
exists at most one bound state; without loss of generality we can assume the existence of exactly one bound state (of non-zero binding energy). Consequently, denoting by ${\bar {\rm {\bf q}}}_{\alpha}$ (${\bar {\rm {\bf q}}}_{\alpha}'$) the incoming (outgoing) on-shell relative momentum between the two fragments in channel $\alpha$, and by $- B_{\alpha} < 0$ the binding 
energy of the bound pair ($\beta \gamma$), energy conservation requires 
\begin{equation} 
E = {\bar q}_{\alpha}^{2}/{2 M_{\alpha}} - B_{\alpha}, \quad {\bar q}_{\alpha}' = {\bar q}_{\alpha}.\label{oes}
\end{equation}

As has already been discussed in I, the only singularity of the effective free Green's function ${\cal G}_{0;\alpha} ({\rm {\bf{q}}}_{\alpha};z)$ is a pole at the on-shell point (\ref{oes}). Hence, 
it remains to investigate the analytical properties of the diagonal 
effective potentials ${\cal V}_{\alpha \alpha}({\rm {\bf{q}}}_{\alpha}',
{\rm {\bf{q}}}_{\alpha};z)$. 
Use of Eqs.\ (\ref{resolv}) leads to 
the following representation 
\begin{eqnarray}
{\cal V}_{\alpha \alpha}({\rm {\bf{q}}}_{\alpha}', 
{\rm {\bf{q}}}_{\alpha};z)= \langle {\rm {\bf{q}}}_{\alpha}', 
\chi_{\alpha} \mid G^{C}_{\alpha}[{\bar V}^{C}_{\alpha} 
+ {\bar V}^{C}_{\alpha}G^{C}{\bar V}^{C}_{\alpha}] G^{C}_{\alpha} \mid 
\chi_{\alpha},{\rm {\bf{q}}}_{\alpha} \rangle, \label{d2efpot1} 
\end{eqnarray}
where the channel Coulomb resolvents $G_{\alpha}^{C}$ have been singled out explicitly. The latter describe the propagation of the
three particles $\alpha$, $\beta$, and $\gamma$, with allowance for 
Coulomb rescattering to all orders between particles 
$\beta$ and $\gamma$ after the virtual decay 
$(\beta \gamma) \to \beta + \gamma$ of the initial bound state
$(\beta \gamma)$, and before the virtual recombination 
$\beta + \gamma \to (\beta \gamma)$ leading to the formation 
of the final bound state $(\beta \gamma)$. The advantage of
doing so arises from the special r\^ole played by 
the Coulomb interactions in the initial and final three-ray 
vertices. Introducing the Coulomb-modified form factor 
$\mid \phi_{\alpha}\rangle := G_{0}^{-1} G_{\alpha}^{C} \mid \chi_{\alpha}\rangle $, see Part I, ${\cal V}_{\alpha \alpha}({\rm {\bf{q}}}_{\alpha}', 
{\rm {\bf{q}}}_{\alpha};z)$ can be rewritten as 
\begin{eqnarray}
{\cal V}_{\alpha \alpha}
({\rm {\bf{q}}}_{\alpha}', {\rm {\bf{q}}}_{\alpha};z)
= {\cal V}^{(a)}_{\alpha \alpha}
({\rm {\bf{q}}}_{\alpha}', {\rm {\bf{q}}}_{\alpha};z) +
{\tilde {\cal V}}_{\alpha \alpha}
({\rm {\bf{q}}}_{\alpha}', {\rm {\bf{q}}}_{\alpha};z),
\label{defpot11}
\end{eqnarray}
with
\begin{eqnarray}
{\cal V}^{(a)}_{\alpha \alpha}({\rm {\bf{q}}}_{\alpha}', 
{\rm {\bf{q}}}_{\alpha};z) 
&=& \langle {\rm {\bf{q}}}_{\alpha}', 
\phi_{\alpha}({\hat z}_{\alpha}'^*) \mid G_{0}(z) {\bar V}_{\alpha}^{C} G_{0}(z) \mid 
\phi_{\alpha}({\hat z}_{\alpha} ),{\rm {\bf{q}}}_{\alpha} \rangle, 
\label{defpot12} \\
{\tilde {\cal V}}_{\alpha \alpha}
({\rm {\bf{q}}}_{\alpha}', {\rm {\bf{q}}}_{\alpha};z) 
&=& \langle {\rm {\bf{q}}}_{\alpha}', 
\phi_{\alpha}({\hat z}_{\alpha}'^*) \mid G_{0}(z) {\bar V}^{C}_{\alpha}
G^{C}(z){\bar V}^{C}_{\alpha} G_{0}(z) \mid 
\phi_{\alpha}({\hat z}_{\alpha} ),{\rm {\bf{q}}}_{\alpha} \rangle. 
\label{defpot13}
\end{eqnarray}
Here, 
\begin{mathletters}
\label{suben}
\begin{eqnarray}
 {\hat z}_{\alpha} &\equiv& {\hat E}_{\alpha} +i0, \quad \mbox{with}\quad {\hat E}_{\alpha} := E - q_{\alpha}^{2}/ 2 M_{\alpha},\label{suben1} \\ 
{\hat z}_{\alpha}' &\equiv& {\hat E}_{\alpha}' +i0, \quad \mbox{with}\quad {\hat E}_{\alpha}' := E - q_{\alpha}'^{2}/ 2 M_{\alpha}, \label{suben2} 
\end{eqnarray}
\end{mathletters}
are the kinetic energies of subsystem $(\beta + \gamma)$ in the initial and final state, respectively.

The first term
${\cal V}^{(a)}_{\alpha \alpha}({\rm {\bf{q}}}_{\alpha}', 
{\rm {\bf{q}}}_{\alpha};z)$ describes elastic scattering
of the projectile particle $\alpha$ off the bound state $(\beta \gamma)$, 
taking into account Coulomb rescattering to all orders 
between particles $\beta$ and $\gamma$ in the initial vertex
$(\beta \gamma) \to \beta + \gamma$ and in the final
vertex $\beta + \gamma \to (\beta \gamma)$ and the single 
intermediate-state Coulomb interaction ${\bar V}_{\alpha}^{C} = V^{C}_{\beta} + V^{C}_{\gamma} $ between particle $\alpha$ and each of the target particles
$\beta$ and $\gamma$. 
The second term ${\tilde {\cal V}}_{\alpha \alpha}
({\rm {\bf{q}}}_{\alpha}', {\rm {\bf{q}}}_{\alpha};z)$ 
differs from that by allowing in the intermediate 
state for Coulomb rescatterings to all orders and between all 
three particles as represented by the three-body Coulomb resolvent. 

An important simplification of the potential part (\ref{defpot12}) occurs at the on-shell point (\ref{oes}). To see this we write down ${\cal V}^{(a)}_{\alpha \alpha}({\rm {\bf{q}}}_{\alpha}', {\rm {\bf{q}}}_{\alpha};z)$ in explicit notation: \begin{eqnarray}
{\cal V}^{(a)}_{\alpha \alpha}({\rm {\bf q}}_{\alpha}', 
{\rm {\bf q}}_{\alpha};z) 
= \frac{4\, \pi\, e_{\alpha}}
{({\rm {\bf q}}_{\alpha}' - {\rm {\bf q}}_{\alpha})^{2}} \; \sum_{\nu=\beta,\gamma} e_{\nu} I^{(\nu)}({\rm {\bf{q}}}_{\alpha}', {\rm {\bf{q}}}_{\alpha};z), 
\label{deffab}
\end{eqnarray}
with, e.g., 
\begin{eqnarray}
I^{(\gamma)}({\rm {\bf{q}}}_{\alpha}', {\rm {\bf{q}}}_{\alpha};z) &=& 4\mu_{\alpha}^{2}\,\langle {\rm {\bf{q}}}_{\alpha}', 
\phi_{\alpha}({\hat z}_{\alpha}'^*) \mid G_{0}(z) G_{0}(z) \mid 
\phi_{\alpha}({\hat z}_{\alpha} ),{\rm {\bf{q}}}_{\alpha} \rangle \nonumber \\
&=& 4\mu_{\alpha}^{2}\,\int \frac{d {\rm {\bf k}}}{(2\pi)^{3}} 
\frac{\phi_{\alpha}^*({\rm{\bf k}}_{\alpha}'; 
{\hat z}_{\alpha}'^{*})\,
\phi_{\alpha}({\rm{\bf k}}_{\alpha}; {\hat z}_{\alpha})}
{[k_{\alpha}'^{2} - 2 \mu_{\alpha}{\hat z}_{\alpha}']\, 
[k_{\alpha}^{2} - 2 \mu_{\alpha}{\hat z}_{\alpha}]}. 
\label{deffptbt2}
\end{eqnarray}
Here, ${\rm {\bf k}}_{\alpha} = \epsilon_{\alpha \beta }
({\rm {\bf k}} + \lambda_{\beta \gamma}{\rm {\bf q}}_{\alpha})$
and ${\rm {\bf k}}_{\alpha}' = \epsilon_{\alpha \beta }
({\rm {\bf k}} + \lambda_{\beta \gamma}{\rm {\bf q}}_{\alpha}')$, 
with $\lambda_{\mu \nu} := m_{\mu}/(m_{\mu} + m_{\nu}),\; \mu \neq \nu$, and $\epsilon_{\alpha \beta }
$ being a sign factor (see Part I). The corresponding expression for $I^{(\beta)}({\rm {\bf{q}}}_{\alpha}', {\rm {\bf{q}}}_{\alpha};z) $ follows from that of $I^{(\gamma)}({\rm {\bf{q}}}_{\alpha}', {\rm {\bf{q}}}_{\alpha};z) $ by interchanging the indices $\beta$ and $\gamma$ in the definition of the momenta ${\rm {\bf k}}_{\alpha} $ and ${\rm {\bf k}}_{\alpha}'$. 

The physical interpretation of $\sum_{\nu=\beta,\gamma} e_{\nu} I^{(\nu)}({\rm {\bf{q}}}_{\alpha}', {\rm {\bf{q}}}_{\alpha};z) $ is that of an off-shell extension of the body form factor of the bound state $(\beta \gamma)$. Indeed, taking into account that on the energy shell the Coulomb-modified form factor $|\phi_{\alpha}({\hat z}_{\alpha} )\rangle$ is related to the bound state wave function $|\psi_{\alpha}\rangle$ via (recall Eqs.\ (I.22) and (I.54))
\begin{equation}
G_{0}({\bar q}_{\alpha}^{2}/{2 M_{\alpha}} - B_{\alpha} +i0) \mid \phi_{\alpha}(-{B}_{\alpha} ), {\bar {\rm {\bf q}}}_{\alpha} \rangle = \mid \psi_{\alpha} , {\bar {\rm {\bf q}}}_{\alpha} \rangle, 
\end{equation}
expression (\ref{deffptbt2}) simplifies to (with ${\tilde{\rm{\bbox{\Delta}}}}_{\alpha} = \epsilon_{\alpha \beta }
\lambda_{\beta \gamma}({\bar {\rm {\bf{q}}}}_{\alpha}' - {\bar {\rm {\bf{q}}}}_{\alpha} )$)
\begin{eqnarray}
I^{(\gamma)}({\bar {\rm {\bf{q}}}}_{\alpha}', {\bar {\rm {\bf{q}}}}_{\alpha} ;E+i0) &=& 
\int \frac{d {\rm {\bf k}}}{(2\pi)^{3}} \,
\psi_{\alpha}^*({\rm{\bf k}} + {\tilde{\rm{\bbox{\Delta}}}}_{\alpha})\,
\psi_{\alpha}({\rm{\bf k}}) \nonumber \\
&=& \int d {\rm {\bf r}} \, e^{i {\tilde{\rm{\bbox{\Delta}}}}_{\alpha} \cdot {\rm {\bf r}} } |{\tilde \psi}_{\alpha}({\rm {\bf r}} )|^2. 
\end{eqnarray}
Thus, assuming the bound state wave function to be normalized to unity, one has on the energy shell in the forward-scattering direction (i.e., for ${\bar {\rm {\bf q}}}_{\alpha} = {\bar {\rm {\bf q}}}_{\alpha}' $)
\begin{eqnarray}
I^{(\nu)}({\bar {\rm {\bf q}}}_{\alpha}, 
{\bar {\rm {\bf q}}}_{\alpha};E+i0) = 1, \quad \nu =\beta,\gamma.\label{iboes}
\end{eqnarray}

\newpage
\subsection{Leading singularity of ${\cal V}_{\alpha \alpha}
({\rm {\bf q}}_{\alpha}', {\rm {\bf q}}_{\alpha};z)$} 
\protect\label{ssba}

Let us state the assertion in the form of a \\
{\bf Theorem:} \\
{\it 
The leading (dynamic) singularity of the diagonal effective potential contribution (\ref{defpot12}) with respect to the momentum transfer is the pole at the border of the physical region, namely at
\begin{eqnarray}
\bbox{\Delta_{\alpha}} := {\rm {\bf q}}_{\alpha}' - {\rm {\bf q}}_{\alpha} = 0,
\label{delt0}
\end{eqnarray}
already displayed in the representation (\ref{deffab}).
Writing
\begin{equation}
{\cal V}^{(a)}_{\alpha \alpha}({\rm {\bf{q}}}_{\alpha}', 
{\rm {\bf{q}}}_{\alpha};z) \;{=:}\; \frac{{\tilde I}({\rm {\bf{q}}}_{\alpha}', {\rm {\bf{q}}}_{\alpha};z)}{\Delta_{\alpha}^{2}} , \label{result1}
\end{equation}
the leading singularity of ${\tilde I}({\rm {\bf{q}}}_{\alpha}', {\rm {\bf{q}}}_{\alpha};z)$ is generally located in the unphysical region and, hence, is harmless. The only exception occurs for ${\hat z}_{\alpha}'>0$ and ${\hat z}_{\alpha}>0$ when it lies in the physical region and is of the type ($a$ is a positive definite constant depending on the particle masses only)
\begin{eqnarray}
{\tilde I}({\rm {\bf{q}}}_{\alpha}', {\rm {\bf{q}}}_{\alpha};z)
&\sim& \left \lbrack a \Delta_{\alpha}^{2} - \left(\sqrt{2 \mu_{\alpha} {\hat z}_{\alpha}'} + \sqrt{2 \mu_{\alpha} {\hat z}_{\alpha}} \right)^{2} \right \rbrack^{\Lambda}, \quad {\hat z}_{\alpha}'>0,{\hat z}_{\alpha}>0 , \label{result2}
\end{eqnarray}
with
\begin{equation}
\Lambda = e_{\beta}e_{\gamma}\mu_{\alpha} / \left(\sqrt{2 \mu_{\alpha} 
{\hat z}_{\alpha}} + \sqrt{2 \mu_{\alpha} {\hat z}_{\alpha}'} \right).\label{result3}
\end{equation}
For particles with charges of equal sign, as they are considered exclusively in the present investigation, such a singularity is, however, not dangerous (but it would give rise to severe problems for $e_{\beta}e_{\gamma} < 0$).

The leading (dynamic) singularity of the effective potential part (\ref{defpot13}) is, for all values of energy, of the form
\begin{equation}
{\tilde {\cal V}}_{\alpha \alpha}
({\rm {\bf{q}}}_{\alpha}', {\rm {\bf{q}}}_{\alpha};z) \stackrel{\Delta_{\alpha} \to 0}{\sim} \frac{1}{\Delta_{\alpha}}.\label{result4}
\end{equation}
}

\newpage
\subsection{Leading singularity of the kernel ${\cal K}_{\alpha \alpha}
({\rm {\bf q}}_{\alpha}', {\rm {\bf q}}_{\alpha};E+i0)$ and its treatment} \protect\label{lskernel}

Given the leading singularity of ${\cal V}_{\alpha \alpha}({\rm {\bf q}}_{\alpha}', {\rm {\bf q}}_{\alpha};E+i0)$, the singularity structure of 
the kernel ${\cal K}_{\alpha \alpha} ({\rm {\bf q}}_{\alpha}',
{\rm {\bf q}}_{\alpha};E+i0)$, Eq.\ (\ref{dkernel}), follows in a straightforward manner. Integration over the right-hand 
variable, presently denoted by ${\rm {\bf q}}_{\alpha}$, is 
implied in (\ref{ags1}); ${\rm {\bf q}}_{\alpha}'$ is a 
vector-valued parameter. The leading 
singularities of the kernel are the pole that originates from
${\cal V}_{\alpha \alpha}({\rm {\bf q}}_{\alpha}',
{\rm {\bf q}}_{\alpha};E+i0)$ and is located as described in 
Sect.\ \ref{ssba}, and the pole of the effective propagator 
\begin{equation}
{\cal G}_{0;\alpha} ({\rm {\bf q}}_{\alpha};z)=
\frac{S_{\alpha}(z- q_{\alpha}^{2}/2M_{\alpha})}
{z- q_{\alpha}^{2}/2M_{\alpha}+ B_{\alpha}}, 
\label{grf1}
\end{equation}
which occurs for $z=E+i0$ at the `on-shell point' $q_{\alpha} = {\bar q}_{\alpha}$. 
(The numerator function $S_{\alpha}(z- q_{\alpha}^{2}/2M_{\alpha})$ is nonsingular, cf.\ I, Appendix B). It is the possibility of the coincidence of these two singularities which renders the diagonal kernel noncompact.

However, this noncompact singularity can be extracted and inverted explicitly, as has been proposed by Veselova \cite{ves70} for energies below the breakup threshold and by Alt and Sandhas \cite{as80,as96} for all energies (within the screening and renormalisation approach). This procedure will now be briefly sketched.

First we recall that on the energy shell (\ref{oes}) one has for normalized bound state wave functions (cf.\ Eq.\ (I.23))
\begin{equation}
S_{\alpha}(- B_{\alpha}) = 1, \quad \mbox{for}\quad \alpha = 1,2,3.\label{soes}
\end{equation}
Using this property it proves convenient to redefine the transition amplitudes as
\begin{eqnarray}
{\cal T}_{\beta \alpha}'({\rm {\bf q}}_{\beta}', 
{\rm {\bf q}}_{\alpha};z) := S_{\beta}^{1/2}(z- q_{\beta}'^{2}/2M_{\beta}) {\cal T}_{\beta \alpha}({\rm {\bf q}}_{\beta}', {\rm {\bf q}}_{\alpha};z) S_{\alpha}^{1/2}(z- q_{\alpha}^{2}/2M_{\alpha}), 
\end{eqnarray}
and similarly for ${\cal V}_{\beta \alpha}'({\rm {\bf q}}_{\beta}', {\rm {\bf q}}_{\alpha};z)$ etc. On account of (\ref{soes}) they coincide on the energy shell with the original quantities, i.e., ${\cal T}_{\beta \alpha}'({\bar {\rm {\bf q}}}_{\beta}', {\bar {\rm {\bf q}}}_{\alpha};E+i0) \equiv {\cal T}_{\beta \alpha}({\bar {\rm {\bf q}}}_{\beta}', {\bar {\rm {\bf q}}}_{\alpha};E+i0) $, etc. 

Furthermore, define the operators
\begin{eqnarray}
{\frak g}_{0;\alpha}({\hat z}) &:=& \left( {\hat z}- Q_{\alpha}^{2}/2M_{\alpha} \right)^{-1}, \label{fg0} \\
{\frak g}_{\alpha}^C({\hat z}) &:=& \left( {\hat z}- Q_{\alpha}^{2}/2M_{\alpha} - {\frak v}_{\alpha}^C \right)^{-1}, \label{fgc}\\
{\frak t}_{\alpha}^C({\hat z}) &=& {\frak v}_{\alpha}^C + {\frak v}_{\alpha}^C {\frak g}_{0;\alpha}({\hat z}) {\frak t}_{\alpha}^C({\hat z}) \nonumber \\
&=& {\frak v}_{\alpha}^C + {\frak t}_{\alpha}^C({\hat z}) {\frak g}_{0;\alpha}({\hat z}) {\frak v}_{\alpha}^C.\label{ftc}
\end{eqnarray}
They all act nontrivially only in the space spanned by the plane waves $|{\rm {\bf q}}_{\alpha}\rangle$ which are eigenstates of the relative momentum operator ${\rm {\bf Q}}_{\alpha}$ between particle $\alpha$ and the center of mass of the subsystem $(\beta + \gamma)$ (cf.\ Part I). For instance, the momentum space representation of ${\frak v}_{\alpha}^C$ is
\begin{equation}
{\frak v}_{\alpha }^C
({\rm {\bf{q}}}_{\alpha}', {\rm {\bf{q}}}_{\alpha}) := \langle {\rm {\bf{q}}}_{\alpha}'|{\frak v}_{\alpha}^C| {\rm {\bf{q}}}_{\alpha} \rangle = \frac{4 \pi e_{\alpha}(e_{\beta} + e_{\gamma})}{ ({\rm {\bf{q}}}_{\alpha}' - {\rm {\bf{q}}}_{\alpha}) ^2}.\label{vcms}
\end{equation}
The physical interpretation of these quantities is evident: ${\frak v}_{\alpha }^C ({\rm {\bf{q}}}_{\alpha}', {\rm {\bf{q}}}_{\alpha})$, which in coordinate space reads as ${\frak v}_{\alpha }^C(\rho_{\alpha}) = e_{\alpha}(e_{\beta} + e_{\gamma})/\rho_{\alpha}$ ($\bbox{\rho}_{\alpha}$ is the coordinate canonically conjugate to ${\rm {\bf{q}}}_{\alpha}$), describes the Coulomb interaction of particle $\alpha$ with a fictitious point particle of charge $(e_{\beta} + e_{\gamma})$ and mass $(m_{\beta} + m_{\gamma})$, and is conventionally called center-of-mass Coulomb potential. ${\frak g}_{0;\alpha}({\hat z})$ is the free propagator of particle $\alpha$ and this fictitious particle, ${\frak g}_{\alpha}^C({\hat z})$ the corresponding propagator with allowance for Coulomb scattering between these two bodies and, finally, ${\frak t}_{\alpha}^C({\hat z})$ the appropriate two-body Coulomb transition operator.

We can now apply, e.g., the procedure detailed in \cite{as80}. We first rewrite Eq.\ (\ref{ags1}) in terms of the `primed' quantities using an operator notation as
\begin{eqnarray}
{\cal T}_{\beta \alpha}'(z) = {\cal V}_{\beta \alpha}'(z) 
+ \sum_{\nu=1}^3 \,{\cal K}_{\beta \nu}' (z) \, {\cal T}_{\nu \alpha}'(z),\label{ags1'} 
\end{eqnarray}
with
\begin{equation}
{\cal K}_{\beta \alpha}' (z) := {\cal V}_{\beta \alpha}' (z) {\frak g}_{0,\alpha}(z + B_{\alpha}). 
\end{equation}
Making use of the Theorem, the total effective potential can be decomposed into a long-ranged (${\frak v}_{\alpha}^{C}$) and a shorter-ranged part as
\begin{equation}
{\cal V}_{\beta \alpha}'(z) = \delta_{\beta \alpha} {\frak v}_{\alpha}^{C} +
{\cal V}_{\beta \alpha}'^{SC}(z).  \label{vsplit}
\end{equation}
The so-called Coulomb-modified short-range effective potential ${\cal V}_{\beta \alpha}'^{SC}(z) $ is given as
\begin{eqnarray}
{\cal V}_{\beta \alpha}'^{SC}(z) := \delta_{\beta \alpha} {\cal V}_{\alpha \alpha}'^{SC}(z) + {\bar \delta}_{\beta \alpha} {\cal V}_{\beta \alpha}'(z).
\end{eqnarray}
Its nondiagonal part coincides with the original effective potential ${\cal V}_{\beta \alpha}'(z), \; \beta \neq \alpha,$ while the diagonal part is defined as
\begin{eqnarray}
{\cal V}_{\alpha \alpha}'^{SC}
({\rm {\bf{q}}}_{\alpha}', {\rm {\bf{q}}}_{\alpha};z)&:=& \frac{4\, \pi\, e_{\alpha}}
{ \Delta_{\alpha}^{2}} \; \sum_{\nu=\beta,\gamma} e_{\nu} \left[ I'^{(\nu)}({\rm {\bf{q}}}_{\alpha}', {\rm {\bf{q}}}_{\alpha};z) - 1 \right] \nonumber \\
&& + {\tilde {\cal V}}_{\alpha \alpha}'
({\rm {\bf{q}}}_{\alpha}', {\rm {\bf{q}}}_{\alpha};z). \label{vcmsr}
\end{eqnarray}
Note that, on account of Eq.\ (\ref{iboes}), each term $[I'^{(\nu)}({\rm {\bf{q}}}_{\alpha}', {\rm {\bf{q}}}_{\alpha};z) - 1], \, \nu = \beta,\gamma,$ vanishes in the forward-scattering on-shell limit ${\rm {\bf q}}_{\alpha}' = {\rm {\bf q}}_{\alpha} = {\bar{\rm {\bf q}}}_{\alpha} $ (where the pole of the propagator ${\frak g}_{0,\alpha}({\rm {\bf{q}}}_{\alpha}; E + i0 + B_{\alpha}) = 2 M_{\alpha} / ({\bar q}_{\alpha}^{2} - q_{\alpha}^{2} + i0 )$ is located). In coordinate space this entails that ${\cal V}_{\alpha \alpha}'^{SC} (\bbox{\rho}_{\alpha}', \bbox{\rho}_{\alpha};z)$ decays asymptotically faster than ${\frak v}_{\alpha }^C(\rho_{\alpha})$. 
Introducing this splitting into (\ref{ags1'}) and 
applying the two-potential procedure leads to the following representation 
\begin{eqnarray}
{\cal T}_{\beta \alpha}'(z) &=& \delta_{\beta \alpha} {\frak t}_{\alpha}^C(z + B_{\alpha}) + \omega_{\beta}^{\dagger}(z^* + B_{\beta})\, {\cal T}_{\beta \alpha}'^{SC}(z) \, \omega_{\alpha}(z + B_{\alpha}).\label{tctsc}
\end{eqnarray}
Here,
\begin{equation}
\omega_{\alpha}^C(z + B_{\alpha}) := [1 + {\frak g}_{0; \alpha}(z + B_{\alpha}){\frak t}_{\alpha}^C(z + B_{\alpha})] \label{moller}
\end{equation}
is the (stationary) off-shell center-of-mass Coulomb M$\O$LLER operator, 
and ${\cal T}_{\beta \alpha}'^{SC}(z)$ is solution of the Lippmann-Schwinger-type equation
\begin{eqnarray}
{\cal T}_{\beta \alpha}'^{SC}(z) = {\cal V}_{\beta \alpha}'^{SC}(z) + \sum_{\nu=1}^3 \,{\cal K}_{\beta \nu}'^{SC}(z) \, {\cal T}_{\nu \alpha}'^{SC}(z), \label{ags3} 
\end{eqnarray}
with kernel
\begin{equation}
{\cal K}_{\beta \alpha}'^{SC}(z) := \,{\cal V}_{\beta \alpha}'^{SC}(z)\, {\frak g}_{\alpha}^C(z + B_{\alpha}).\label{ksc}
\end{equation}

As has already been pointed out, for $\beta \neq \alpha$ we have ${\cal V}_{\beta \alpha}'^{SC}({\rm {\bf q}}_{\beta}', {\rm {\bf q}}_{\alpha};z) \equiv {\cal V}_{\beta \alpha}'({\rm {\bf q}}_{\beta}',{\rm {\bf q}}_{\alpha};z)$, which has been shown in Part I to possess no nonintegrable singularities. Its diagonal part ${\cal V}_{\alpha\alpha}'^{SC}({\rm {\bf q}}_{\alpha}', {\rm {\bf q}}_{\alpha};z)$, on the other hand, differs from the original effective potential ${\cal V}_{\alpha \alpha}'({\rm {\bf q}}_{\alpha}',{\rm {\bf q}}_{\alpha};z) $ by the absence in the former of the compactness destroying center-of-mass Coulomb potential ${\frak v}_{\alpha}^C({\rm {\bf{q}}}_{\alpha}',{\rm {\bf{q}}}_{\alpha}) $; 
thus, its leading singular behavior is of the type (\ref{result4}). 
The result is that the leading singularities of the kernel ${\cal K}_{\beta \alpha}'^{SC}({\rm {\bf q}}_{\beta}', {\rm {\bf q}}_{\alpha};z)$ of the modified equation (\ref{ags3}) are integrable, with the consequence that ${\cal K}_{\beta \alpha}'^{SC}({\rm {\bf q}}_{\beta}', {\rm {\bf q}}_{\alpha};z)$ becomes compact after a suitable number of iterations. 
This result verifies the assertion made in \cite{as80}.

Once Eq.\ (\ref{ags3}) has been solved for the amplitudes ${\cal T}_{\beta \alpha}'^{SC}({\rm {\bf q}}_{\beta}', {\rm {\bf q}}_{\alpha}; z)$ by standard solution methods (see below), the physical on-shell arrangement amplitudes ${\cal T}_{\beta \alpha}({\bar {\rm {\bf q}}}_{\beta}', {\bar {\rm {\bf q}}}_{\alpha})$ are easily obtained. For, sandwiching the operator relation (\ref{tctsc}) between plane waves $\langle {\rm {\bf{q}}}_{\beta}'|$ and $| {\rm {\bf{q}}}_{\alpha} \rangle$ and applying the standard on-shell limiting procedures for two-body Coulombian quantities (see, e.g., Refs.\ \cite{haer85,a86}), yields the following representation 
\begin{eqnarray}
{\cal T}_{\beta \alpha}({\bar {\rm {\bf q}}}_{\beta}', 
{\bar {\rm {\bf q}}}_{\alpha}) := \delta_{\beta \alpha} {\frak t}_{\alpha}^C({\bar {\rm {\bf q}}}_{\alpha}', {\bar {\rm {\bf q}}}_{\alpha}) + \langle {\bar {\rm {\bf q}}}_{\beta }'^{C(-)} | {\cal T}_{\beta \alpha}'^{SC}(E+i0) | {\bar {\rm {\bf q}}}_{\alpha}^{C(+)} \rangle .\label{tctscoes}
\end{eqnarray}
The quantity ${\frak t}_{\alpha}^C({\bar {\rm {\bf q}}}_{\alpha}, 
{\bar {\rm {\bf q}}}_{\alpha}) $ is the two-body Rutherford amplitude, describing Coulomb scattering of particle $\alpha$ off the total charge of particles $\beta$ and $\gamma$ concentrated in their center of mass, and $| {\bar {\rm {\bf q}}}_{\alpha}^{C(\pm)} \rangle $ are the corresponding center-of-mass Coulomb scattering states, both of which are explicitly known. Practical evaluation of the second term, called Coulomb-modified short-range transition operator in the `Coulomb representation', which comprises all effects coming from the shorter-ranged Coulomb effective potential parts and from the genuine short-range interactions, is straightforward (see below). The full amplitude ${\cal T}_{\beta \alpha}({\bar {\rm {\bf q}}}_{\beta}', {\bar {\rm {\bf q}}}_{\alpha})$, however, as the notation indicates can not be obtained as solution of some compact-kernel integral equation but only via Eq.\ (\ref{tctscoes}) (the same holds true also for ${\frak t}_{\alpha}^C({\bar {\rm {\bf q}}}_{\alpha}, {\bar {\rm {\bf q}}}_{\alpha}) )$. 

For completeness we mention that the definition (\ref{tctscoes}) of the physical charged-composite particle amplitudes agrees with that following from the time-dependent scattering theory \cite{as78} and from the stationary screening and renormalization approach \cite{as80}.

\newpage
\subsection{Practical approaches} \protect\label{practical}

As described above, the full, on-shell, charged-particle reaction amplitudes ${\cal T}_{\beta \alpha}({\bar {\rm {\bf q}}}_{\beta}', {\bar {\rm {\bf q}}}_{\alpha}) $, Eq.\ (\ref{tctscoes}), can not be obtained as solutions of some integral equations by standard methods (due to the noncompactness of the kernels (\ref{dkernel})). The same situation, in fact, arises for the on-shell center-of-mass Coulomb amplitude ${\frak t}_{\alpha}^C({\bar {\rm {\bf q}}}_{\alpha}', {\bar {\rm {\bf q}}}_{\alpha}) $. Instead, one first has to calculate the (on-shell) Coulomb-modified short-range transition amplitudes in the Coulomb representation, $\langle {\bar {\rm {\bf q}}}_{\beta }'^{C(-)} | {\cal T}_{\beta \alpha}'^{SC}(E+i0) | {\bar {\rm {\bf q}}}_{\alpha}^{C(+)} \rangle $, and then to add according to (\ref{tctscoes}) the analytically known center-of-mass Coulomb amplitude. 
To reach this goal two strategies have been developed so far.

\subsubsection{Screening and renormalisation approach}

Development of this approach followed the analogous development in stationary two-charged particle scattering \cite{t74}. 
The basic idea is to use screened Coulomb potentials,
\begin{equation}
V_{\alpha}^R(r) = V_{\alpha}^C(r) g^{R}(r), \label{vscr} 
\end{equation}
where $g^{R}(r)$ is some fairly arbitrary but smooth screening function with
\begin{eqnarray}
\lim_{R \to \infty} g^{R}(r) &=& 1 \quad (r \quad \mbox{fixed}), \\
\lim_{r \to \infty} g^{R}(r) &=& 0 \quad (R \quad \mbox{fixed}).
\end{eqnarray}
A numerically convenient form is $g^{R}(r) = \exp\{-r/R\}$. Consequently, all three-body quantities will depend on the screening radius R. Since, for finite $R$, the potentials (\ref{vscr}) are of short range, standard methods of integral equations theory are applicable to the equation for the screened arrangement amplitudes:
\begin{eqnarray}
{\cal T}_{\beta \alpha}^{(R)}({\rm {\bf q}}_{\beta}', 
{\rm {\bf q}}_{\alpha};z) = {\cal V}_{\beta \alpha}^{(R)}
({\rm {\bf q}}_{\beta}', {\rm {\bf q}}_{\alpha};z) 
+ \sum_{\nu=1}^3\;\int {\frac{d{\rm {\bf q}}''_{\nu}}{({2 \pi})^{3}}} 
\,{\cal K}_{\beta \nu}^{(R)}({\rm {\bf q}}_{\beta}', 
{\rm {\bf q}}_{\nu}'';z) 
\, {\cal T}_{\nu \alpha}^{(R)}({\rm {\bf q}}''_{\nu}, \,
{\rm {\bf q}}_{\alpha};z), \label{ags1scr} 
\end{eqnarray}
as the kernels
\begin{eqnarray} 
{\cal K}_{\beta \alpha}^{(R)}({\rm {\bf{q}}}_{\beta }', 
{\rm {\bf{q}}}_{\alpha};z) := {\cal V}_{\beta \alpha}^{(R)}
({\rm {\bf{q}}}_{\beta }', {\rm {\bf{q}}}_{\alpha};z) 
{\cal G}_{0;\alpha}^{(R)}({\rm {\bf{q}}}_{\alpha};z) \label{kernelscr}
\end{eqnarray}
become compact after a suitable number of iterations.

It remains to recover the desired unscreened amplitudes (\ref{tctscoes}) by a well-defined limiting procedure $R \to \infty$. Indeed, it has been proven in \cite{asz78,as80} that after multiplication of the screened on-shell arrangement amplitudes ${\cal T}_{\beta \alpha}^{(R)}({\bar {\rm {\bf q}}}_{\beta}', {\bar {\rm {\bf q}}}_{\alpha};E+i0)$ with suitable renormalisation factors $Z_{\beta,R}^{-1/2}({\bar q}_{\beta}')$ and $Z_{\alpha,R}^{-1/2}({\bar q}_{\alpha}) $ which for large $R$ are fully determined by the choice of the screening function $g^{R}(r)$, 
the following limits exist 
\begin{eqnarray}
\lim_{R \to \infty} Z_{\beta,R}^{-1/2}({\bar q}_{\beta}') {\cal T}_{\beta \alpha}^{(R)}({\bar {\rm {\bf q}}}_{\beta}', {\bar {\rm {\bf q}}}_{\alpha};E+i0) Z_{\alpha,R}^{-1/2}({\bar q}_{\alpha}) = {\cal T}_{\beta \alpha}({\bar {\rm {\bf q}}}_{\beta}', {\bar {\rm {\bf q}}}_{\alpha}) , \label{tctscoes1}
\end{eqnarray}
and yield the desired unscreened amplitudes (\ref{tctscoes}). This is the approach used in the various numerical applications (for a list of references see \cite{as96,ams98} where also the full particulars of how to proceed in practice can be found).

\subsubsection{Direct solution of Eq.\ \protect(\ref{ags3})}

An alternative strategy \cite{as80} which aims at directly calculating the Coulomb-modified short-range amplitude in the `Coulomb representation' $\langle {\bar {\rm {\bf q}}}_{\beta }'^{C(-)} | {\cal T}_{\beta \alpha}'^{SC}(E+i0) | {\bar {\rm {\bf q}}}_{\alpha}^{C(+)} \rangle $, is based on Eq.\ (\ref{ags3}). Indeed, sandwiching this equation between Coulomb scattering states $\langle {\rm {\bf q}}_{\beta }'^{C(-)}| $ and $|{\rm {\bf q}}_{\alpha}^{C(+)} \rangle $, and using in the kernel (\ref{ksc}) the spectral representation of the resolvent ${\frak g}_{\nu}^C({\hat z}_{\nu})$ in the form (recall that all Coulomb potentials are assumed repulsive)
\begin{equation}
{\frak g}_{\nu}^C({\hat z}_{\nu}) = \int \frac{d^3 q_{\nu}''}{(2\pi)^3}\; \frac{| {\rm {\bf q}}_{\nu}''^{C(-)} \rangle \langle {\rm {\bf q}}_{\nu}''^{C(-)} | }{{\hat z}_{\nu} - q_{\nu}''^{2}/2 M_{\nu}},
\end{equation}
we end up with
\begin{eqnarray}
\langle {\rm {\bf q}}_{\beta}'^{C(-)}| {\cal T}_{\beta \alpha}'^{SC}(z)| {\rm {\bf q}}_{\alpha}^{C(+)} \rangle &=& \langle {\rm {\bf q}}_{\beta}'^{C(-)}| {\cal V}_{\beta \alpha}'^{SC}(z)| {\rm {\bf q}}_{\alpha}^{C(+)} \rangle + \nonumber \\
&& \sum_{\nu=1}^3 \int \frac{d^3 q_{\nu}''}{(2\pi)^3}\; \frac{\langle {\rm {\bf q}}_{\beta}'^{C(-)}| {\cal V}_{\beta \nu}'^{SC}(z)| {\rm {\bf q}}_{\nu}''^{C(-)} \rangle  \langle {\rm {\bf q}}_{\nu}''^{C(-)} | {\cal T}_{\nu \alpha}'^{SC}(z)| {\rm {\bf q}}_{\alpha}^{C(+)} \rangle}{z - q_{\nu}''^{2}/2 M_{\nu} + B_{\nu} } .
\end{eqnarray}
As input one has to provide the effective potentials in the Coulomb representation, $\langle {\rm {\bf q}}_{\beta}'^{C(-)}| {\cal V}_{\beta \alpha}'^{SC}(z)| {\rm {\bf q}}_{\alpha}^{C(\pm)} \rangle$, the calculation of which, however, appears feasible at best in coordinate space (although for two-particle scattering a momentum space calculation along these lines has been performed successfully in \cite{dpsw79}).

\newpage
\section{Proofs of the assertions} \protect\label{koff}

\subsection{Leading singularity of 
${\cal V}^{(a)}_{\alpha \alpha}({\rm {\bf q}}_{\alpha}', 
{\rm {\bf q}}_{\alpha};z)$}

From the explicit representation (\ref{deffab}) it is seen that 
in the $\Delta_{\alpha}^{2}-$plane ${\cal V}^{(a)}_{\alpha \alpha}({\rm {\bf{q}}}_{\alpha}', {\rm {\bf{q}}}_{\alpha};z)$ has the familiar Coulomb forward-scattering singularity at $\Delta_{\alpha}^{2}= 0$ originating from the Fourier transform of the Coulomb channel interaction ${\bar V}_{\alpha}^{C}$. However, additional singularities arise from the integral terms $I^{(\beta)}({\rm {\bf{q}}}_{\alpha}', {\rm {\bf{q}}}_{\alpha};z)$ and $I^{(\gamma)}({\rm {\bf{q}}}_{\alpha}', {\rm {\bf{q}}}_{\alpha};z)$, cf.\ Eq.\ (\ref{deffptbt2}).

1) We begin our investigation of the singular behavior
of $I^{(\gamma)}({\rm {\bf{q}}}_{\alpha}', {\rm {\bf{q}}}_{\alpha};z) $ in the momentum transfer plane, for momenta
\begin{equation}
q_{\alpha},q_{\alpha}' \not= {\tilde q}_{\alpha} := \sqrt{2 M_{\alpha} E} 
\label{qaatilde}
\end{equation}
which is equivalent to (cf.\ Eq.\ (\ref{suben}))
\begin{equation}
{\hat E}_{\alpha} \not= 0 \quad \mbox{and}\quad {\hat E}_{\alpha}' \not=0. 
\end{equation}
In that case the singular behavior of the off-shell Coulomb-modified form factors is as explicated in Eq.\ (I.62a) (this implies, of course, certain analyticity requirements for the nuclear form factors $\chi_{\alpha}({\rm{\bf k}}_{\alpha})$, cf.\ I, Appendix C). Thus, expression (\ref{deffptbt2}) can be rewritten as 
\begin{eqnarray}
I^{(\gamma)}({\rm {\bf{q}}}_{\alpha}', {\rm {\bf{q}}}_{\alpha};z) = 4\mu_{\alpha}^{2}\, \int \frac{d {\rm {\bf k}}}{(2\pi)^{3}} 
\frac{{\tilde \phi}_{\alpha}^*({\rm{\bf k}}_{\alpha}'; 
{\hat z}_{\alpha}'^{*})\,
{\tilde \phi}_{\alpha}({\rm{\bf k}}_{\alpha}; 
{\hat z}_{\alpha})} {[k_{\alpha}'^{2} - 
2 \mu_{\alpha}{\hat z}_{\alpha}']^{1 - i{\hat \eta}_{\alpha}'}\, 
[k_{\alpha}^{2} - 
2 \mu_{\alpha}{\hat z}_{\alpha}]^{1 - i{\hat \eta}_{\alpha}}}. 
\label{J0}
\end{eqnarray}
The Coulomb parameters are defined as
\begin{equation}
{\hat \eta}_{\alpha} = \frac{e_{\beta}e_{\gamma}\mu_{\alpha}}
{\sqrt{2\mu_{\alpha}{\hat z}_{\alpha}}},
\quad
{\hat \eta}_{\alpha}' = \frac{e_{\beta}e_{\gamma}\mu_{\alpha}}
{\sqrt{2\mu_{\alpha}{\hat z}_{\alpha}'}}. 
\label{tdeta}
\end{equation}

The leading singularity of $I^{(\gamma)}({\rm {\bf{q}}}_{\alpha}', {\rm {\bf{q}}}_{\alpha};z) $ is generated by the coincidence
of the zeroes of the denominator at 
\begin{equation}
k_{\alpha}^{2} - 2 \mu_{\alpha}{\hat z}_{\alpha} = 0 \label{fsing}
\end{equation}
and at
\begin{equation}
k_{\alpha}'^{2} - 2 \mu_{\alpha}{\hat z}_{\alpha}' = 0.\label{ssing}
\end{equation}

To simplify the derivation 
we assume without loss of generality that the orbital angular momentum of the relative motion of particles $\beta$ and $\gamma$ in the initial and final bound states $(\beta \gamma)$ is zero. As a consequence, the reduced Coulomb-modified form factors ${\tilde \phi}_{\alpha}^*({\rm{\bf k}}_{\alpha}'; {\hat z}_{\alpha}'^{*})$ and ${\tilde \phi}_{\alpha}({\rm{\bf k}}_{\alpha}; {\hat z}_{\alpha})$ which are regular functions at (\ref{ssing}) and (\ref{fsing}), respectively, can be taken out from under the integral sign at these points (otherwise only their radial parts could be taken out). 
Thus, in the leading order we end up with
\begin{eqnarray}
I^{(\gamma)}({\rm {\bf{q}}}_{\alpha}', {\rm {\bf{q}}}_{\alpha};z) \sim 4\mu_{\alpha}^{2}\, {\tilde \phi}_{\alpha}^{*}\left(\sqrt{2 \mu_{\alpha} {\hat z}_{\alpha}'^{*}} ; {\hat z}_{\alpha}'^{*}\right) \, 
{\tilde \phi}_{\alpha}\left(\sqrt{2 \mu_{\alpha} {\hat z}_{\alpha}}; {\hat z}_{\alpha}\right)\,{\tilde I}^{(\gamma)}({\rm {\bf{q}}}_{\alpha}', {\rm {\bf{q}}}_{\alpha};z), 
\end{eqnarray}
where
\begin{eqnarray}
{\tilde I}^{(\gamma)}({\rm {\bf{q}}}_{\alpha}', {\rm {\bf{q}}}_{\alpha};z)&=& \int \frac{d {\rm {\bf k}}}{(2\pi)^{3}} \frac{1}
{[({\rm {\bf k}} + \lambda_{\beta \gamma}{\rm {\bf q}}_{\alpha}')^{2} - 
2 \mu_{\alpha}{\hat z}_{\alpha}']^{1 - i{\hat \eta}_{\alpha}'} }
\nonumber \\
&&\times 
\frac{1}
{[({\rm {\bf k}} + \lambda_{\beta \gamma}{\rm {\bf q}}_{\alpha})^{2} - 
2 \mu_{\alpha}{\hat z}_{\alpha}]^{1 - i{\hat \eta}_{\alpha}}}. 
\label{dft1}
\end{eqnarray}

To find the singular behavior of ${\tilde I}^{(\gamma)}$ we make use of the method described in \cite{am94}. It employs the intimate and unique connection 
between the singularity of a function which is nearest to the physical region
in the $y$-plane, where $y = {\rm {\bf {\hat q}}}_{\alpha}' \cdot
{\rm {\bf {\hat q}}}_{\alpha}$ is the cosine of the scattering angle,
and the behavior of its partial wave projections for $\ell \to \infty$
\cite{pop64}. If the singularity lies outside the physical region defined by $-1 \leq y \leq +1$, as it happens in the present case, application of this method proceeds in a straightforward manner.

For illustration consider the function $a^{(\pm)}(y) = 1/(\zeta - y)^{1 \pm i\eta}$, with $\zeta \not \in [-1,1]$. The partial wave expansion \cite{dol66,am94} 
\begin{equation}
\frac{1}{(\zeta - y)^{1 \pm i\eta}}  
= \sum\limits_{\ell=0}^{\infty}
(2\ell + 1)\,P_{\ell}(y)\,a_{\ell}^{(\pm)}(\zeta), \; y \in [-1,1],\label{prtex1}
\end{equation}
defines the partial wave projections $a_{\ell}^{(\pm)}(\zeta)$ as
\begin{equation}
a_{\ell}^{(\pm)}(\zeta) = \frac{1}{2}\int\limits_{-1}^{+1}\,
{\rm d}y \frac{P_{\ell}(y)}{(\zeta - y)^{1 \pm i\eta}}=
\frac{i}{2\pi}\,(1 - e^{\pm 2\pi\eta})\, \Gamma(\mp i\eta)\,
(\zeta^{2} - 1)^{\mp i\eta/2}\,Q_{\ell}^{\pm i\eta}(\zeta).
\label{prtin1}
\end{equation}
Here, $P_{\ell}(y)$ are the Legendre polynomials and 
$Q_{\ell}^{\lambda}(\zeta)$ the associated Legendre functions 
of the second kind. 

For the following we require the asymptotic formulae
\cite{KrFr63} 
\begin{equation}
Q_{\ell}^{\lambda}(\zeta) \stackrel{\ell \to \infty}{\approx}
e^{i\pi \lambda}\,\ell^{\lambda}\,Q_{\ell}(\zeta), \label{qas1}
\end{equation}
\begin{equation}
Q_{\ell}(\zeta) \stackrel{\ell \to \infty}{=}
\sqrt{\frac{\pi}{\ell}}\frac{e^{-\ell\ln{\tau}}}{\sqrt{\tau^{2} -1}} +
o\left(\frac{1}{\sqrt{\ell}}\right), \label{qas2}
\end{equation}
with 
\begin{equation}
\tau \equiv \tau ( \zeta) =\zeta + \sqrt{\zeta^{2} - 1}.\label{ta1}
\end{equation}
Note that our assumption $\zeta \not \in [-1,1]$ implies $\tau \neq 1$. 

The behavior of $a_{\ell}^{(\pm)}(\zeta)$ for $\ell \to \infty$ now follows immediately as
\begin{eqnarray}
a_{\ell}^{(\pm)}(\zeta) 
\stackrel{\ell \to \infty}{\approx} \pm
\frac{\sqrt{\pi}\,(\zeta^{2}-1)^{\mp i \eta/2}\,\ell^{\pm i \eta -1/2}}
{\Gamma(1 \pm i \eta)}\,
\frac{e^{-\ell \ln{\tau}}}{\sqrt{\tau^{2} - 1}}, 
\label{apart1}
\end{eqnarray}
where use has been made of
\begin{equation}
e^{\pm \pi \eta} (1 - e^{\mp 2 \pi \eta})\, 
\Gamma(\pm i \eta)= \frac{\mp 2 i \pi}
{\Gamma(1 \mp i \eta)}.\label{etag1}
\end{equation}

Thus, we have the result that partial wave 
amplitudes $a_{\ell}^{(\pm)}(\zeta)$, which behave asymptotically
(i.e., for $\ell \to \infty$) as in (\ref{apart1}),
generate a singularity $\sim 1/(\zeta - y)^{1 \pm i\eta}$, cf.\ Eq.\ (\ref{prtex1}).
The location $\zeta$ of the singularity can, for instance, be read off from 
$\tau(\zeta)$ as $\zeta =(\tau + \tau^{-1})/2$, and the singularity strength factor $\eta$ from the corresponding exponent of $\ell$. 

Let us apply this result to ${\tilde I}^{(\gamma)}({\rm {\bf{q}}}_{\alpha}', 
{\rm {\bf{q}}}_{\alpha};z) $, Eq.\ (\ref{dft1}). We introduce the notation
\begin{equation}
\zeta = \zeta(k) = \frac{k^{2} + \lambda_{\beta \gamma}^{2}
q_{\alpha}^{2} - 2 \mu_{\alpha}{\hat z}_{\alpha}}
{2 \lambda_{\beta \gamma} k q_{\alpha}}, \quad 
\zeta' = \zeta'(k) = \frac{k^{2} + 
\lambda_{\beta \gamma}^{2}{q_{\alpha}'}^{2} - 
2 \mu_{\alpha}{\hat z}_{\alpha}'}
{2 \lambda_{\beta \gamma} k q_{\alpha }'}, \label{zt1}
\end{equation} 
and consider first the case ${\hat z}_{\alpha} \equiv {\hat E}_{\alpha} <0$, 
${\hat z}_{\alpha}' \equiv {\hat E}_{\alpha}' <0$, 
which yields $\zeta > 1$ and $\zeta' > 1$. 
When performing a partial wave expansion of ${\tilde I}^{(\gamma)}({\rm {\bf q}}_{\alpha}', {\rm {\bf q}}_{\alpha};z) $,
\begin{equation}
{\tilde I}^{(\gamma)}({\rm {\bf q}}_{\alpha}', {\rm {\bf q}}_{\alpha};z) = 
\sum \limits_{\ell=0}^{\infty}
(2\ell + 1)\,P_{\ell}(y) {\tilde I}_{\ell}^{(\gamma)}, \quad y= {\rm {\bf {\hat q}}}_{\alpha}' \cdot 
{\rm {\bf {\hat q}}}_{\alpha}, \label{prtexv}
\end{equation}
the following expression for the expansion coefficients
is obtained, 
\begin{eqnarray}
{\tilde I}_{\ell}^{(\gamma)}(q_{\alpha}', q_{\alpha};z)
&=& -\frac{(2\lambda_{\beta \gamma})^{-2+ i({\hat \eta}_{\alpha}'+ {\hat \eta}_{\alpha})}}{8\pi^{4}}\,(1 - e^{-2\pi {\hat \eta}_{\alpha}'})\,
(1 - e^{-2\pi {\hat \eta}_{\alpha}})\, 
\Gamma(i{\hat \eta}_{\alpha}')\,
\Gamma(i{\hat \eta}_{\alpha})\,
{q_{\alpha}'^{-1 + i{\hat \eta}_{\alpha}'}}
{q_{\alpha}^{-1 + i{\hat \eta}_{\alpha}}} \nonumber \\ 
&& \times 
\int\limits_{0}^{\infty} {\rm d} k \,k^{i({\hat \eta}_{\alpha}' + {\hat \eta}_{\alpha})}\,
({\zeta'}^{2} - 1)^{i{\hat \eta}_{\alpha}'/2}\, 
({\zeta}^{2} - 1)^{i{\hat \eta}_{\alpha}/2} 
Q_{\ell}^{-i{\hat \eta}_{\alpha}'}(\zeta')
Q_{\ell}^{-i{\hat \eta}_{\alpha}}(\zeta). 
\label{dpt1}
\end{eqnarray}

Define $\tau = \tau(\zeta)$ as in (\ref{ta1}), and similarly 
$\tau'$ with $\zeta$ replaced by $\zeta'$. Note that $\tau \neq 1$ ($\tau' \neq 1$) as $\zeta \neq 1$ ($\zeta' \neq 1$). 
Use of Eqs.\ (\ref{qas1}) and (\ref{qas2}) gives for the large-$\ell$ behavior
\begin{eqnarray} 
{\tilde I}_{\ell}^{(\gamma)}(q_{\alpha}', q_{\alpha};z)
&\stackrel{\ell \to \infty}{\approx}& 
\frac{ (2 \lambda_{\beta \gamma})^{-2+ i({\hat \eta}_{\alpha}'+ {\hat \eta}_{\alpha})}}{2\pi}
\frac{q_{\alpha}'^{-1 + i{\hat \eta}_{\alpha}'}\,
q_{\alpha}^{- 1 + i{\hat \eta}_{\alpha}}}
{\Gamma(1-i{\hat \eta}_{\alpha}')\,
\Gamma(1-i{\hat \eta}_{\alpha})} \; 
\ell^{- i({\hat \eta}_{\alpha}' + {\hat \eta}_{\alpha}) - 1} 
\nonumber \\ 
&& \times \int\limits_{0}^{\infty} 
{\rm d} k \, k^{i({\hat \eta}_{\alpha}' +
{\hat \eta}_{\alpha})}\,
({\zeta'}^{2} - 1)^{i{\hat \eta}_{\alpha}'/2}\, 
({\zeta}^{2} - 1)^{i{\hat \eta}_{\alpha}/2}\, 
\frac{e^{-\ell\ln{\tilde \tau}}}
{\sqrt{({\tau'}^{2} - 1)(\tau^{2} - 1)}}, \label{dpt2}
\end{eqnarray}
with ${\tilde \tau} := {\tau}' \, {\tau}$. 

The remaining integral in (\ref{dpt2}) is evaluated by means of the saddle point method. As the whole $\ell$-dependence of the integrand resides in the exponential, for $\ell \to \infty$ the main contribution to the integral comes
from the region around the saddle point. The latter can be found by solving
the equation
\begin{equation}
\frac{{\rm d} \ln{\tilde \tau}(k)}{{\rm d}k} = 0,\label{drsp1}
\end{equation}
which determines the minimum of the function $\ln{\tilde \tau}(k)$. 
Straighforward algebra yields for the location of the saddle point 
\begin{equation}
k = k_{\rm (sp)} := \left[\frac{\lambda_{\beta \gamma}^{2} \left(q_{\alpha}'^{2} \sqrt{-{\hat z}_{\alpha}}
+ q_{\alpha}^{2} \sqrt{{-\hat z}_{\alpha}'}\right)} 
{ \sqrt{{-\hat z}_{\alpha}}+\sqrt{-{\hat z}_{\alpha}'}} + 
2 \mu_{\alpha} \sqrt{-{\hat z}_{\alpha}}\sqrt{-{\hat z}_{\alpha}'}\right]^{1/2}. \label{sdp1}
\end{equation}

When calculating the contribution from the saddle point to the integral in Eq.\ (\ref{dpt2}), all factors of the integrand 
which are nonsingular at $k = k_{\rm (sp)}$ and slowly
varying in the neighbourhood of $k_{\rm (sp)}$, 
can be taken out from under the integral sign at $k = k_{\rm (sp)}$. 
Then we immediately arrive at 
\begin{eqnarray} 
{\tilde I}_{\ell}^{(\gamma)}(q_{\alpha}', q_{\alpha};z)
&\stackrel{\ell \to \infty}{\approx}& 
\frac{1}{\sqrt{2\pi}}\,
\frac{(2\lambda_{\beta \gamma})^{-2+ i({\hat \eta}_{\alpha}'+ {\hat \eta}_{\alpha})}}{\Gamma(1 - i{\hat \eta}_{\alpha}')\,
\Gamma(1 - i{\hat \eta}_{\alpha})} \,
\frac{q_{\alpha}'^{-1 + i{\hat \eta}_{\alpha}'}\,
q_{\alpha}^{-1 + i{\hat \eta}_{\alpha}}} 
{\sqrt{\left(\ln {\tilde \tau} \right)''\mid_{k=k_{\rm (sp)}}}} 
\nonumber \\ 
&& \times k_{\rm (sp)}^{i({\hat \eta}_{\alpha} +
{\hat \eta}_{\alpha}')}\,
\left(\zeta_{\rm (sp)}'^{2}) - 1 \right)^{i{\hat \eta}_{\alpha}'/2}\, 
\left(\zeta_{\rm (sp)}^{2}) - 1 \right)^{i{\hat \eta}_{\alpha}/2}
\nonumber \\
&& \times \ell^{ - i({\hat \eta}_{\alpha}' + {\hat \eta}_{\alpha}) - 3/2}\,
 \frac{e^{-\ell\ln{\tilde \tau}(k_{\rm (sp)})}}
{\sqrt{\left({\tau'}^{2}(\zeta_{\rm (sp)}') - 1 \right)\left(\tau^{2}(\zeta_{\rm (sp)}) - 1 \right)}}\;, 
\label{dpt2'}
\end{eqnarray}
where we have introduced the notation
\begin{equation}
\zeta_{\rm (sp)} := \zeta(k_{\rm (sp)}), \quad \zeta_{\rm (sp)}' := \zeta'(k_{\rm (sp)}).\label{zksp}
\end{equation}
Double prime means second derivative.

Let us define ${\tilde \zeta}$ in terms of ${\tilde \tau}$ as in (\ref{ta1}), or explicitly 
\begin{equation}
{\tilde \zeta} = ({\tilde \tau}+ {\tilde \tau}^{-1})/2. 
\label{tildt1}
\end{equation}
It then follows from Eqs. (\ref{tildt1})
and (\ref{sdp1}) that 
\begin{equation}
{\tilde \zeta}_{\rm (sp)} := {\tilde \zeta}(k_{\rm (sp)}) = 
\frac{q_{\alpha}^{2} + {q_{\alpha}'}^{2} +
\left(\sqrt{-2 \mu_{\alpha}{\hat z}_{\alpha}'} + 
\sqrt{-2 \mu_{\alpha}{\hat z}_{\alpha}}\right)^{2} /\lambda_{\beta \gamma}^{2}}
{2\,q_{\alpha}q_{\alpha}'}. 
\label{tildz1}
\end{equation}

Given this asymptotic behavior (for $\ell \to \infty$) 
of ${\tilde I}_{\ell}^{(\gamma)}(q_{\alpha}', q_{\alpha};z)$, by comparison with (\ref{apart1}) 
and (\ref{prtex1}) we immediately recover the leading singularity of
${\tilde I}^{(\gamma)}({\rm {\bf{q}}}_{\alpha}', {\rm {\bf{q}}}_{\alpha};z)$ as
\begin{eqnarray}
{\tilde I}^{(\gamma)}({\rm {\bf{q}}}_{\alpha}', {\rm {\bf{q}}}_{\alpha};z)
&\sim& \frac{1}{({\tilde \zeta}_{sp} - y)^{- i({\hat \eta}_{\alpha}' + 
{\hat \eta}_{\alpha})}} \label{jsing1}\\
&\sim& {\left \lbrack \lambda_{\beta \gamma}^{2} ({\rm {\bf q}}_{\alpha}' - 
{\rm {\bf q}}_{\alpha})^{2} + \left(\sqrt{-2 \mu_{\alpha}{\hat z}_{\alpha}'} + 
\sqrt{-2 \mu_{\alpha}{\hat z}_{\alpha}}\right)^{2} 
\right \rbrack^{ i({\hat \eta}_{\alpha}' + 
{\hat \eta}_{\alpha})}}.\label{jsing2}
\end{eqnarray}

It is not difficult to see that the derivation goes through unaltered if $\zeta'$ and/or $\zeta $ have a nonvanishing imaginary part 
as it happens if ${\hat E}_{\alpha}' > 0$ and/or 
${\hat E}_{\alpha} > 0$. 


From (\ref{jsing2}) it follows that the leading singularity of ${\tilde I}^{(\gamma)}({\rm {\bf{q}}}_{\alpha}', {\rm {\bf{q}}}_{\alpha};z)$
never coincides with the Coulomb forward-scattering singularity $\Delta_{\alpha}^{2}=0$ 
(recall that presently we assume 
${\hat z}_{\alpha}' \not= 0$ and 
${\hat z}_{\alpha} \not= 0$). 
In fact, if both ${\hat z}_{\alpha} < 0$ and 
${\hat z}_{\alpha}' < 0$, 
it lies farther away from the physical region
than the forward Coulomb singularity and hence is not dangerous. 
A similar situation arises if ${\hat z}_{\alpha}'$ or 
${\hat z}_{\alpha} $ is positive, since then the singularity
of ${\tilde I}^{(\gamma)}({\rm {\bf{q}}}_{\alpha}', {\rm {\bf{q}}}_{\alpha};z)$
is located in the complex $\Delta_{\alpha}^{2}-$plane. However, if 
both ${\hat z}_{\alpha}'$ and ${\hat z}_{\alpha} $ are positive, 
${\tilde I}^{(\gamma)}({\rm {\bf{q}}}_{\alpha}', {\rm {\bf{q}}}_{\alpha};z)$ becomes 
singular in the physical region. 
Indeed, for $z = E + i0$ we have 
$\sqrt{-{\hat z}_{\alpha}} = + i \sqrt{{\hat z}_{\alpha}}$, 
and similarly for $\sqrt{-{\hat z}_{\alpha}'}$. Consequently, 
\begin{eqnarray}
{\tilde I}^{(\gamma)}({\rm {\bf{q}}}_{\alpha}', {\rm {\bf{q}}}_{\alpha};z)
&\sim& \left \lbrack \lambda_{\beta \gamma}^{2} \Delta_{\alpha}^{2} - \left(\sqrt{2 \mu_{\alpha} {\hat z}_{\alpha}'} + 
\sqrt{2 \mu_{\alpha} {\hat z}_{\alpha}} \right)^{2} \right \rbrack^{\Lambda}, \quad {\hat z}_{\alpha}'>0, \, {\hat z}_{\alpha}>0, \label{jsing3}
\end{eqnarray}
with $\Lambda $ given in (\ref{result3}).
This singularity is, however, not dangerous 
if particles $\beta$ and $\gamma$ have charges of the same sign, i.e., if $e_{\beta}e_{\gamma} > 0$. 

It is obvious that the leading singularity of ${\tilde I}^{(\beta)}({\rm {\bf{q}}}_{\alpha}', {\rm {\bf{q}}}_{\alpha};z)$ which is closest to the physical region, coincides with that shown in Eq.\ (\ref{jsing2}), except for the replacement $\lambda_{\beta \gamma} \rightarrow \lambda_{\gamma \beta}$. Hence, the above discussion of its location holds without change also for the present case.

2) It remains to consider the cases ${\hat z}_{\alpha} = 0$ or/and ${\hat z}_{\alpha}' = 0$. To start with assume ${\hat z}_{\alpha} = 0$, i.e.\ $q_{\alpha} = {\tilde q}_{\alpha}$, but ${\hat z}_{\alpha}' \neq 0$. Taking into account the behavior of 
$\phi_{\alpha}({\rm{\bf k}}_{\alpha};0)$ in the limit $k_{\alpha} \to 0$,
as described in Eq.\ (I.62b), we have instead of expression (\ref{J0}) 
\begin{eqnarray}
I^{(\gamma)}({\rm {\bf{q}}}_{\alpha}', {\tilde {\rm {\bf q}}}_{\alpha} ;z) = 
4\mu_{\alpha}^{2}\,\int \frac{d {\rm {\bf k}}}{(2\pi)^{3}} 
\frac{{\tilde \phi}_{\alpha}^{*}({\rm{\bf k}}_{\alpha}'; 
{\hat z}_{\alpha}'^{*})\,
{\tilde \phi}_{\alpha}({\rm{\bf k}}_{\alpha}; 0)}
{[{k_{\alpha}'}^{2} - 
2 \mu_{\alpha}{\hat z}_{\alpha}']^{1 - i{\hat \eta}_{\alpha}'}}. 
\label{dfpp1}
\end{eqnarray}
Here, ${\rm{\bf k}}_{\alpha}'$ and ${\rm{\bf k}}_{\alpha}$ are defined as before but with ${\rm {\bf q}}_{\alpha}$ replaced by ${\tilde {\rm {\bf q}}}_{\alpha}:= {\hat{\rm{\bf q}}}_{\alpha} {\tilde q}_{\alpha}$.
The leading singular behavior of the integral 
in the $\Delta_{\alpha}^{2}-$plane is generated by the coincidence 
of the zero of the denominator of the integrand with the closest singularity of 
${\tilde \phi}_{\alpha}({\rm{\bf k}}_{\alpha}; 0)$. The location
of the latter depends on the large-distance properties of the short-range (nuclear) potential 
between particles $\beta$ and $\gamma$. For instance, for a Yukawa-type form factor $\chi_{\alpha}(r)\sim r^{-1} \exp(-\nu_{\alpha} r)$, where $1/{\nu_{\alpha}}$
measures the range of the nuclear interaction, 
the closest singularity of ${\tilde \phi}_{\alpha}({\rm{\bf k}}_{\alpha}; 0)$ lies at 
\begin{equation}
k_{\alpha}^{2} + \nu_{\alpha}^{2}=0 \label{yukp1}
\end{equation}
(cf.\ Eq.\ (I.C.15)). Hence, the singularity of 
$I^{(\gamma)}({\rm {\bf{q}}}_{\alpha}', 
{\tilde {\rm {\bf q}}}_{\alpha};z)$, which is generated by the coincidence 
of singularities at (\ref{ssing}) and (\ref{yukp1}), is 
located at 
\begin{equation}
\lambda_{\beta \gamma}^{2} \Delta_{\alpha}^{2}=
- \left(\sqrt{-2 \mu_{\alpha}{\hat z}_{\alpha}'} + \nu_{\alpha}\right)^{2}.\label{sng1}
\end{equation}
A similar situation occurs for ${\hat z}_{\alpha}' = 0$, but ${\hat z}_{\alpha} \neq 0$. Finally, for ${\hat z}_{\alpha}' = {\hat z}_{\alpha} = 0$ simultaneously, the expression for 
$I^{(\gamma)}({\tilde {\rm {\bf q}}}_{\alpha}', 
{\tilde {\rm {\bf q}}}_{\alpha};z)$ (${\tilde {\rm {\bf q}}}_{\alpha}':= {\hat{\rm{\bf q}}}_{\alpha}'{\tilde q}_{\alpha}$) reads as
\begin{eqnarray}
I^{(\gamma)}({\tilde {\rm {\bf q}}}_{\alpha}', 
{\tilde {\rm {\bf q}}}_{\alpha};z) = 4\mu_{\alpha}^{2}\,
\int \frac{d {\rm {\bf k}}}{(2\pi)^{3}} 
{\tilde \phi}_{\alpha}^*({\rm{\bf k}}_{\alpha}';0)\,
{\tilde \phi}_{\alpha}({\rm{\bf k}}_{\alpha}; 0), 
\label{dfpp2}
\end{eqnarray}
with ${\rm{\bf k}}_{\alpha}$ expressed by ${\tilde {\rm {\bf q}}}_{\alpha}$ and ${\rm{\bf k}}_{\alpha}'$ by ${\tilde {\rm {\bf q}}}_{\alpha}':= {\hat{\rm{\bf q}}}_{\alpha}' {\tilde q}_{\alpha}$.
In this case the leading singularity of the integral in the
$\Delta_{\alpha}^{2}-$plane is generated by 
the coincidence of the singularities of 
${\tilde \phi}_{\alpha}^*({\rm{\bf k}}_{\alpha}';0)$
and ${\tilde \phi}_{\alpha}({\rm{\bf k}}_{\alpha}; 0)$
and is, for Yukawa-type form factors, located at
\begin{equation}
\lambda_{\beta \gamma}^{2} \Delta_{\alpha}^{2}=
- (2\nu_{\alpha})^{2}.\label{sng2}
\end{equation}
The singularities at (\ref{sng1}) and (\ref{sng2}) always lie outside the physical region, and hence are not dangerous. 

The behavior of $I^{(\beta)}({\rm {\bf{q}}}_{\alpha}', {\rm {\bf{q}}}_{\alpha};z)$ follows again from the above by the substitution $\lambda_{\beta \gamma}\rightarrow \lambda_{\gamma \beta}$.


3) Taken together we have thus shown that in the leading order the singular behavior of
${\cal V}^{(a )}_{\alpha \alpha}({\rm {\bf{q}}}_{\alpha}', 
{\rm {\bf{q}}}_{\alpha};z)$ is of the asserted form, namely, 
\begin{eqnarray}
{\cal V}^{(a)}_{\alpha \alpha}({\rm {\bf{q}}}_{\alpha}', 
{\rm {\bf{q}}}_{\alpha};z) \stackrel{\Delta_{\alpha} \to 0}{\sim} \frac{1}
{\Delta_{\alpha}^{2}} \;
\sum_{\rho \neq \alpha} \left[ \frac{ \mu_{\alpha}}{m_{\rho} } \Delta_{\alpha}^{2}+ u_{\alpha}^{2} \right ]^{i({\hat \eta}_{\alpha}' + 
{\hat \eta}_{\alpha})} + o \left( \frac{1}{\Delta_{\alpha}^{2}} \right)
\label{nua1}
\end{eqnarray}
where, assuming a Yukawa-type behavior for the nuclear form factors with inverse range $\nu_{\alpha} > 0$, 
\begin{eqnarray}
u_{\alpha} = \left \lbrace \begin{array}{l}
\sqrt{-2 \mu_{\alpha} {\hat z}_{\alpha}'} + 
\sqrt{-2 \mu_{\alpha}{\hat z}_{\alpha}} \quad \mbox{for} \quad {\hat z}_{\alpha}' \neq 0,\, {\hat z}_{\alpha} \neq 0,\\ 
\sqrt{-2 \mu_{\alpha} {\hat z}_{\alpha}'} + \nu_{\alpha} \quad \mbox{for} \quad {\hat z}_{\alpha}' \neq 0, \,{\hat z}_{\alpha} = 0,\\ 
2 \nu_{\alpha} \quad \mbox{for} \quad {\hat z}_{\alpha}' = 0 = {\hat z}_{\alpha} .
\end{array}\right. 
\end{eqnarray}

This proves the first part of the Theorem. 

\newpage
\subsection{Leading singularity of 
${\tilde {\cal V}}_{\alpha \alpha}({\rm {\bf q}}_{\alpha}', 
{\rm {\bf q}}_{\alpha};z)$}
\label{ldvaa1}

\subsubsection{Introductory remarks}

Consider now the 
contribution (\ref{defpot13}) to the effective potential which is abbreviated as ${\tilde {\cal V}}_{\alpha \alpha} ({\rm {\bf{q}}}_{\alpha}', {\rm {\bf{q}}}_{\alpha};z)=
\langle {\rm {\bf{q}}}_{\alpha}', 
\phi_{\alpha}\mid {\cal O} \mid \phi_{\alpha},{\rm {\bf{q}}}_{\alpha} \rangle$. The operator ${\cal O}:= G_{0} {\bar V}^{C}_{\alpha}
G^{C}{\bar V}^{C}_{\alpha} G_{0}$ contains 
all possible intermediate-state Coulomb interactions: it begins with the 
incoming-channel Coulomb interaction ${\bar V}^{C}_{\alpha}$, ends up with the outgoing-channel Coulomb interaction ${\bar V}^{C}_{\alpha}$, while the three-body Coulomb resolvent $G^{C}$ takes into account all possible Coulomb rescatterings of particles $\alpha,\, \beta,$ and $\gamma$, in between. If, in $\cal O$, $G^{C}$ is replaced by the free resolvent $G_{0}$, which is the first term in a Neumann series expansion of $G^{C}$, the lowest-order contribution 
\begin{equation}
{\tilde {\cal V}}_{\alpha \alpha}^{(2)}
({\rm {\bf{q}}}_{\alpha}', {\rm {\bf{q}}}_{\alpha};z)
= \langle {\rm {\bf{q}}}_{\alpha}', 
\phi_{\alpha}\mid G_{0} {\bar V}^{C}_{\alpha}
G_{0}{\bar V}^{C}_{\alpha} G_{0} \mid 
\phi_{\alpha},{\rm {\bf{q}}}_{\alpha} \rangle
\label{vaa2}
\end{equation}
results. Below it is shown that the leading singularities of ${\tilde {\cal V}}_{\alpha \alpha}({\rm {\bf{q}}}_{\alpha}', {\rm {\bf{q}}}_{\alpha};z)$ and ${\tilde {\cal V}}_{\alpha \alpha}^{(2)} ({\rm {\bf{q}}}_{\alpha}', {\rm {\bf{q}}}_{\alpha};z)$ coincide. In other words, near the leading singularity 
the three-body Coulomb resolvent $G^{C}$ may effectively be replaced by
the free resolvent $G_{0}$. Note the similarity of this assertion to that encountered for the nondiagonal effective potential in I.

{\bf Auxiliary Theorem}: {\it Even an arbitrary number of Coulomb rescatterings in 
the intermediate state of the diagonal effective potential contribution
${\tilde {\cal V}}_{\alpha \alpha}({\rm {\bf{q}}}_{\alpha}', 
{\rm {\bf{q}}}_{\alpha};z)$, as represented by the three-body 
Coulomb resolvent $G^{C}$, does change neither position 
nor character of the leading (dynamic) singularity in the 
momentum transfer variable as compared to its lowest-order
contribution ${\tilde {\cal V}}_{\alpha \alpha}^{(2)}
({\rm {\bf{q}}}_{\alpha}', {\rm {\bf{q}}}_{\alpha};z)$, 
but does alter the strength of the residue.}

\subsubsection{Leading singularity of 
${\tilde {\cal V}}_{\alpha \alpha}^{(2)}
({\rm {\bf{q}}}_{\alpha}', {\rm {\bf{q}}}_{\alpha};z)$} \label{ss4}

\begin{figure}[htb]
\protect
\epsfig{file=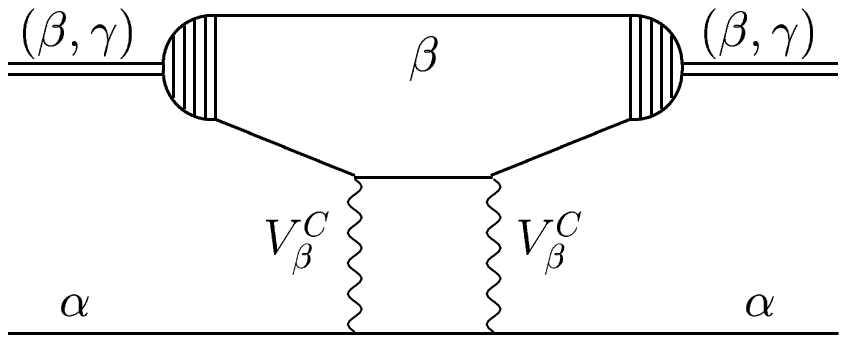,clip=,width=117mm,height=46mm}\\
\caption{Lowest-order contribution (\protect\ref{tildfp20}) to the diagonal effective potential. The dashed semi-circles represent Coulomb-modified form factors.}
\label{fig2}
\end{figure}

We start with the investigation of 
${\tilde {\cal V}}_{\alpha \alpha}^{(2)}
({\rm {\bf{q}}}_{\alpha}', {\rm {\bf{q}}}_{\alpha};z)$. 
According to its definition (\ref{vaa2}), it
can be written as a sum of four terms,
\begin{equation}
{\tilde {\cal V}}_{\alpha \alpha}^{(2)}
({\rm {\bf{q}}}_{\alpha}', {\rm {\bf{q}}}_{\alpha};z)=
\sum_{\nu, \sigma \not= \alpha} 
\langle {\rm {\bf{q}}}_{\alpha}', 
\phi_{\alpha} \mid G_{0} V_{\nu}^{C}G_{0} V^{C}_{\sigma}
G_{0} \mid 
\phi_{\alpha},{\rm {\bf{q}}}_{\alpha} \rangle. 
\label{tildfp10}
\end{equation}
Consider, for example, the term with $\sigma= \beta$ and
$\nu = \beta$, to be denoted as
\begin{equation}
{\tilde {\cal V}}^{(2)(\beta \beta)}_{\alpha \alpha}
({\rm {\bf{q}}}_{\alpha}', {\rm {\bf q}}_{\alpha};z):=
\langle {\rm {\bf q}}_{\alpha}', 
\phi_{\alpha} \mid G_{0}V^{C}_{\beta}G_{0}V^{C}_{\beta}
G_{0} \mid \phi_{\alpha},{\rm {\bf q}}_{\alpha} \rangle
\label{tildfp20}
\end{equation}
and represented in diagrammatic form in Fig.\ \ref{fig2}. 

1) For ${\hat z}_{\alpha}' \not= 0$ and ${\hat z}_{\alpha} \not= 0$, the singular behavior of the Coulomb-modified form factor $\phi_{\alpha}({\rm {\bf k}}_{\alpha}; {\hat z}_{\alpha})$ is as given in Eq.\ (I.62a). Inserting the momentum space representation of the spectral resolution of the free resolvent in the form (I.84) and the explicit expression for the Fourier transform of the Coulomb potential $V^{C}_{\beta}$, we end up with
\begin{eqnarray}
{\tilde {\cal V}}^{(2)(\beta \beta)}_{\alpha \alpha}
({\rm {\bf{q}}}_{\alpha}', {\rm {\bf{q}}}_{\alpha};z) &=&
4 \mu_{\alpha}^{2}
\, \int {\frac{d{\rm {\bf {q}}}_{\beta}^{0}}{({2 \pi})^{3}}} 
\,\int {\frac{d{\rm {\bf {q}}}_{\alpha}^{0}}{({2 \pi})^{3}}}\,
\frac {{\tilde \phi}^{*}_{\alpha}\left(\epsilon_{\alpha \beta} 
({\rm {\bf q}}_{\beta}^{0} + \lambda_{\beta \gamma}
{\rm {\bf {q}}}_{\alpha}')\right)}{\left[({\rm {\bf {q}}}_{\beta}^{0} + 
\lambda_{\beta \gamma}{\rm {\bf {q}}}_{\alpha}')^{2}
 - 2\mu_{\alpha}{\hat z}_{\alpha}')\right]^{1 - i{\hat \eta}_{\alpha}'}}\, 
\nonumber \\
&&\times \frac{4{\pi}e_{\alpha}e_{\gamma}}
{\left({\rm {\bf {q}}}_{\alpha}^{0}- {\rm {\bf {q}}}_{\alpha}' \right)^{2}} 
\frac{1}{\left[z - {({\rm {\bf {q}}}_{\beta}^{0} + 
\lambda_{\beta\gamma}{\rm {\bf {q}}}_{\alpha}^{0})^2}/{2\mu_{\alpha}} - 
{q^{{0}^2}_{\alpha}}/{2M_{\alpha}}\right]} 
\nonumber \\
&&\times \frac{4{\pi}e_{\alpha}e_{\gamma}} {\left({\rm {\bf {q}}}_{\alpha}^{0} - 
{\rm {\bf {q}}}_{\alpha}\right)^{2}} 
\frac {{\tilde \phi}_{\alpha}\left( \epsilon_{\alpha \beta}
({\rm {\bf {q}}}_{\beta}^{0} + 
\lambda_{\beta \gamma}{\rm {\bf {q}}}_{\alpha})\right)}
{\left[ ({\rm {\bf q}}_{\beta}^{0} + 
\lambda_{\beta \gamma}{\rm {\bf q}}_{\alpha})^{2}
- 2\mu_{\alpha}{\hat z}_{\alpha})\right]^{1 - i{\hat \eta}_{\alpha}}}. 
\label{tildint2}
\end{eqnarray}
Here and in the following the arguments ${\hat z}_{\alpha}$ in the reduced Coulomb-modified form factors ${\tilde \phi}_{\alpha}(\cdot\,;{\hat z}_{\alpha})$ are dropped unless required for clarity. 

The leading singularity of ${\tilde {\cal V}}^{(2){(\beta \beta)}}_{\alpha 
\alpha}
({\rm {\bf q}}_{\alpha }', {\rm {\bf q}}_{\alpha};z)$ 
is generated by the coincidence of the singularities at
\begin{eqnarray}
{\mbox {\boldmath $\Delta$}}_{\alpha}^{0} := {\rm {\bf q}}_{\alpha}^{0}- 
{\rm {\bf q}}_{\alpha} = 0, \label{sing04} 
\end{eqnarray}
and at
\begin{eqnarray}
{\mbox {\boldmath $\Delta$}}_{\alpha}'^{0} := {\rm {\bf q}}_{\alpha}^{0}- 
{\rm {\bf q}}_{\alpha}' = 0.\label{sing5}
\end{eqnarray}
The solution of these two equations gives for
the location of the singularity
\begin{equation}
{\mbox {\boldmath $\Delta$}}_{\alpha}= {\rm {\bf q}}_{\alpha}' 
- {\rm {\bf q}}_{\alpha} = 0.\label{dltsng1}
\end{equation}
Let us introduce as new integration variables ${\mbox {\boldmath $\Delta$}}_{\alpha}^{0}$ and ${\rm {\bf k}}_{\alpha}^{0} = \epsilon_{\alpha \beta} \left({\rm {\bf q}}_{\beta}^{0} + \lambda_{\beta \gamma}{\rm {\bf q}}_{\alpha}^{0}\right)$.
Then (\ref{tildint2}) takes the form
\begin{eqnarray}
&&{\tilde {\cal V}}^{(2){(\beta \beta)}}_{\alpha \alpha}
({\rm {\bf{q}}}_{\alpha}', {\rm {\bf{q}}}_{\alpha};z) = 
 (2 \mu_{\alpha})^{3}
\, \int {\frac{d{\rm {\bf k}}_{\alpha}^{0}}{({2 \pi})^{3}}} 
\int {\frac{d{\mbox {\boldmath $\Delta$}}_{\alpha}^{0}}{({2 \pi})^{3}}}
\nonumber \\ 
&& \times \frac {{\tilde \phi}^{*}_{\alpha}
\left({\rm {\bf k}}_{\alpha}^{0} - \epsilon_{\alpha \beta}\,
\lambda_{\beta \gamma}
[{\mbox {\boldmath $\Delta$}}_{\alpha}^{0} - 
{\mbox {\boldmath $\Delta$}}_{\alpha}] \right)}
{\left[ \left({\rm {\bf k}}_{\alpha}^{0} -\epsilon_{\alpha \beta}\,
\lambda_{\beta \gamma}
[{\mbox {\boldmath $\Delta$}}_{\alpha}^{0} - 
{\mbox {\boldmath $\Delta$}}_{\alpha}] \right)^{2} - 
2\mu_{\alpha}{\hat z}_{\alpha} - 
{\mu_{\alpha}}{\mbox {\boldmath $\Delta$}}_{\alpha}\cdot
\left(2\,{\rm{\bf q}}_{\alpha}' - 
{\mbox {\boldmath $\Delta$}}_{\alpha}\right)/{M_{\alpha}} 
\right]^{1 - i{\hat \eta}_{\alpha}'}} 
\nonumber \\
&&\times \frac{4{\pi}e_{\alpha}e_{\gamma}}
{\left({\mbox {\boldmath $\Delta$}}_{\alpha}^{0} - 
{\mbox {\boldmath $\Delta$}}_{\alpha}\right)^{2}}
\frac{1}{\left[ 2\mu_{\alpha}\,{\hat z}_{\alpha} -
{k_{\alpha}^{0}}^{2} - {\mu_{\alpha}}
{\mbox {\boldmath $\Delta$}}_{\alpha}^{0} \cdot
({\mbox {\boldmath $\Delta$}}_{\alpha}^{0} +
2\,{\rm {\bf q}}_{\alpha})/{M_{\alpha}}\right] } \nonumber \\
&&\times \frac{4{\pi}e_{\alpha}e_{\gamma}}{({\Delta_{\alpha}^{0}})^{2}}
\frac {{\tilde \phi}_{\alpha}\left( {\rm {\bf k}}_{\alpha}^{0} - \epsilon_{\alpha \beta}\,\lambda_{\beta \gamma}
{\mbox {\boldmath $\Delta$}}_{\alpha}^{0} \right)}
{\left[ \left({\rm {\bf k}}_{\alpha}^{0} -\epsilon_{\alpha \beta}\,
\lambda_{\beta \gamma}
{\mbox {\boldmath $\Delta$}}_{\alpha}^{0} \right)^{2}
- 2\,\mu_{\alpha}{\hat z}_{\alpha} \right]^{1 - i{\hat \eta}_{\alpha}}}. 
\label{tildint3}
\end{eqnarray}
Here, we have expressed ${\mbox {\boldmath $\Delta$}}_{\alpha}'^{0} $ according to (\ref{sing04}) and (\ref{dltsng1}), and have taken into account that 
${\hat z}_{\alpha}'= z - {q_{\alpha}'^{2}}/{2\,M_{\alpha}}=
{\hat z}_{\alpha} + {{\mbox {\boldmath $\Delta$}}_{\alpha} \cdot
(2\,{\rm {\bf q}}_{\alpha}' - {\mbox {\boldmath $\Delta$}}_{\alpha})}/
{2\,M_{\alpha}}$.

Application of the scaling transformation
\begin{equation}
{\mbox {\boldmath $\Delta$}}_{\alpha}^{0} =
\Delta_{\alpha} \,{\rm {\bf v}}_{\alpha} \label{sctd1}
\end{equation}
yields 
\begin{equation}
{\tilde {\cal V}}^{(2){(\beta \beta)}}_{\alpha \alpha}
({\rm {\bf{q}}}_{\alpha}', {\rm {\bf{q}}}_{\alpha};z) \stackrel
{\Delta_{\alpha} \to 0}{=} 
\frac{(4\,\pi\,e_{\alpha}\,e_{\gamma})^{2}}{\Delta_{\alpha}}\,
J^{(\beta)}({\hat z}_{\alpha}),
\label{vt2ba1}
\end{equation}
where 
\begin{eqnarray}
J^{(\beta)}({\hat z}_{\alpha}) 
&=& -(2 \mu_{\alpha})^{3}
\int \frac{d{\rm {\bf v}}_{\alpha}}{({2 \pi})^{3} \,v_{\alpha}^{2}}\,
\frac{1}{( {\rm {\bf v}}_{\alpha} - 
{\hat {\mbox {\boldmath $\Delta$}}}_{\alpha} )^{2}}\,
\int \frac{d{\rm {\bf k}}_{\alpha}^{0}}{({2 \pi})^{3}}
\frac {\mid
{\tilde \phi}_{\alpha} \left( {\rm {\bf k}}_{\alpha}^{0}\right)\mid^{2}}
{\left[{k_{\alpha}^{0}}^{2} -
2\mu_{\alpha}{\hat z}_{\alpha} \right]^{3 - 2i{\hat \eta}_{\alpha}}} 
\nonumber \\
&=&- \mu_{\alpha}^{3}
\int \frac{d{\rm {\bf k}}_{\alpha}^{0}}{({2 \pi})^{3}}
\frac {\mid
{\tilde \phi}_{\alpha} \left( {\rm {\bf k}}_{\alpha}^{0}\right)\mid^{2}}
{\left[{k_{\alpha}^{0}}^{2} -
2\mu_{\alpha}{\hat z}_{\alpha} \right]^{3 - 2i{\hat \eta}_{\alpha}}} 
\label{v2ba2}
\end{eqnarray}
Because we presently restrict ourselves to the case ${\hat z}_{\alpha} \not= 0$, the integral in $J^{(\beta)}({\hat z}_{\alpha}) $ is nonsingular as, for finite three-body energies, the singularity of the integrand can not coincide with the integration limits.

Let us add two comments.\\
(i) In order to prove that the leading singularity of ${\tilde {\cal V}}^{(2)(\beta \beta)}_{\alpha \alpha}({\rm {\bf{q}}}_{\alpha}', {\rm {\bf{q}}}_{\alpha};z)$ is due to the coincidence of the singularities of the integrand at (\ref{sing04}) and (\ref{sing5}), it was necessary to take into account the singular behavior of the Coulomb-modified form factors $\phi_{\alpha}({\rm {\bf k}}_{\alpha}; {\hat z}_{\alpha})$ 
as given in Eq.\ (I.62a).\\
(ii) We point out that a behavior $\sim 1/\Delta_{\alpha}$ is typical for a second-order Coulombic contribution (cf.\ the analogous result for the second-order term in the iteration of the Lippmann-Schwinger equation for the two-body Coulomb T-matrix derived in \cite{am94}).

2) Next assume $q_{\alpha}= {\tilde q}_{\alpha}$, i.e.\ 
${\hat z}_{\alpha}=0$, in which case relation (I.62b) applies for $\phi_{\alpha}({\rm {\bf k}}; 0)$. Since we are looking for the behavior of 
${\tilde {\cal V}}^{(2){(\beta \beta)}}_{\alpha \alpha}
({\rm {\bf{q}}}_{\alpha}', {\rm {\bf{q}}}_{\alpha};z)$
near the singularity at (\ref{dltsng1}), 
$\Delta_{\alpha}=0$ implies also ${\hat z}_{\alpha}'=0$ or equivalently $q_{\alpha}' = {\tilde q}_{\alpha}$. Hence, taking into account the scaling substitution (\ref{sctd1}) we find
\begin{equation}
\phi^{*}_{\alpha}\left({\rm {\bf k}}_{\alpha}^{0} -
\epsilon_{\alpha \beta}\,\lambda_{\beta \gamma}
({\mbox {\boldmath $\Delta$}}_{\alpha}^{0} - 
{\mbox {\boldmath $\Delta$}}_{\alpha});{\hat z}_{\alpha}'^{*}\right) \stackrel
{\Delta_{\alpha} \to 0}{=} \phi^{*}_{\alpha}
\left({\rm {\bf k}}_{\alpha}^{0}; 0 \right) = {k_{\alpha}^{0}}^{2}\;{\tilde \phi}^{*}_{\alpha}
\left({\rm {\bf k}}_{\alpha}^{0}; 0 \right) .\label{phpz1}
\end{equation}
Thus, the leading singular term of 
${\tilde {\cal V}}^{(2){(\beta \beta)}}_{\alpha \alpha}
({\rm {\bf{q}}}_{\alpha}', {\rm {\bf q}}_{\alpha};z)$
is given by
\begin{equation}
{\tilde {\cal V}}^{(2){(\beta \beta)}}_{\alpha \alpha}
({\rm {\bf{q}}}_{\alpha}', {\rm {\bf q}}_{\alpha};z)
\stackrel{\Delta_{\alpha} \to 0}{=} 
\frac{(4\,\pi\,e_{\alpha}\,e_{\gamma})^{2}}{\Delta_{\alpha}}\,
J^{(\beta)}(0),
\label{vt2ba2}
\end{equation}
with 
\begin{eqnarray}
J^{(\beta)}(0) = -\mu_{\alpha}^{3} \int \frac{d{\rm {\bf k}}_{\alpha}^{0}}{({2 \pi})^{3}} \, \frac{{\tilde \phi}^{*}_{\alpha}\left({\rm {\bf k}}_{\alpha}^{0};0\right)\, {\tilde \phi}_{\alpha} \left( {\rm {\bf k}}_{\alpha}^{0};0\right)} {{k_{\alpha}^{0}}^{2}} 
\label{v2ba3}
\end{eqnarray}
being nonsingular.

3) An analogous argumentation shows that any effective potential contribution
${\tilde {\cal V}}^{(2){(\sigma \nu)}}_{\alpha \alpha}
({\rm {\bf{q}}}_{\alpha}', {\rm {\bf{q}}}_{\alpha};z)$ with
$\sigma, \, \nu \not= \alpha$, has the same leading singularity 
given by Eq.\ (\ref{vt2ba1}) and Eq.\ (\ref{vt2ba2}), respectively.
Summarizing we have shown that 
${\tilde {\cal V}}^{(2)}_{\alpha \alpha}
({\rm {\bf{q}}}_{\alpha}', {\rm {\bf{q}}}_{\alpha};z)$
behaves, for all values of $q_{\alpha}'$ and $q_{\alpha}$, in the limit ${\Delta_{\alpha} \to 0}$ as
\begin{equation}
{\tilde {\cal V}}^{(2)}_{\alpha \alpha}
({\rm {\bf{q}}}_{\alpha}', {\rm {\bf{q}}}_{\alpha};z)
\stackrel{\Delta_{\alpha} \to 0}{\sim} \frac{1}{\Delta_{\alpha}}.\label{v2sn2}
\end{equation}

\subsubsection{Leading singularity of ${\tilde {\cal V}}_{\alpha
\alpha}({\rm {\bf{q}}}_{\alpha}', 
{\rm {\bf{q}}}_{\alpha};z)$} \label{ss5}

\noindent
{\bf Proof of the Auxiliary Theorem:} 
We are now ready to prove that the 
contribution ${\tilde {\cal V}}_{\alpha \alpha}
({\rm {\bf{q}}}_{\alpha}', {\rm {\bf{q}}}_{\alpha};z)$ 
to the full effective potential, in spite of
containing an infinite number of Coulombic rescatterings between all 
three particles in the 
intermediate state as represented by the three-body Coulomb 
resolvent, possesses the same leading singularity at the same position 
(\ref{dltsng1}) as the lowest-order contribution 
${\tilde {\cal V}}^{(2)}_{\alpha \alpha}
({\rm {\bf{q}}}_{\alpha}', {\rm {\bf{q}}}_{\alpha};z)$, 
cf.\ Eq.\ (\ref{v2sn2}). 

\begin{figure}[htb]
\protect
\epsfig{file=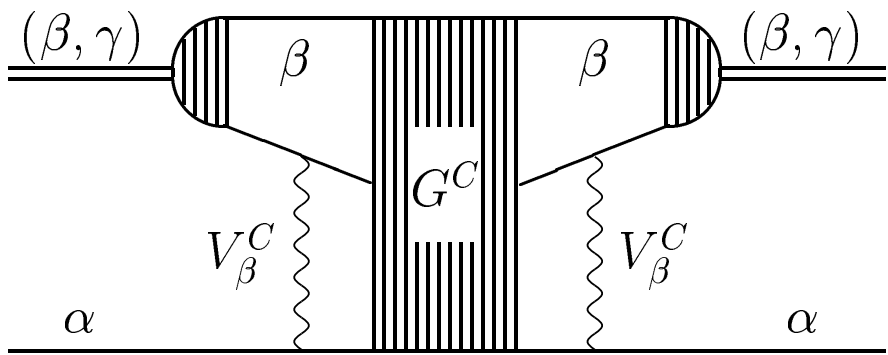,clip=,width=117mm,height=46mm}\\
\caption{Contribution (\protect\ref{tildfp2}) to the exact diagonal effective potential. }
\label{fig1}
\end{figure}

According to its definition (\ref{defpot13}), also
${\tilde {\cal V}}_{\alpha \alpha}
({\rm {\bf{q}}}_{\alpha}', {\rm {\bf{q}}}_{\alpha};z)$
is a sum of four terms,
\begin{eqnarray}
{\tilde {\cal V}}_{\alpha\alpha}
({\rm {\bf{q}}}_{\alpha}', {\rm {\bf{q}}}_{\alpha};z) &=&
\sum_{\nu, \sigma \not= \alpha} 
\langle {\rm {\bf{q}}}_{\alpha}', 
\phi_{\alpha} \mid G_{0} V_{\nu}^{C}G^{C} V^{C}_{\sigma}
G_{0} \mid \phi_{\alpha},{\rm {\bf{q}}}_{\alpha} \rangle \nonumber \\
&=:& \sum_{\nu, \sigma \not= \alpha} 
{\tilde {\cal V}}_{\alpha \alpha}^{(\nu \sigma)}
({\rm {\bf{q}}}_{\alpha}', {\rm {\bf{q}}}_{\alpha};z). 
\label{tildfp1}
\end{eqnarray}
As an example we investigate 
\begin{equation}
{\tilde {\cal V}}^{(\beta \beta)}_{\alpha \alpha}
({\rm {\bf{q}}}_{\alpha}', {\rm {\bf q}}_{\alpha};z):=
\langle {\rm {\bf q}}_{\alpha}', 
\phi_{\alpha} \mid G_{0}V^{C}_{\beta}G^{C}V^{C}_{\beta}
G_{0} \mid \phi_{\alpha},{\rm {\bf q}}_{\alpha} \rangle
\label{tildfp2}
\end{equation}
which is represented in diagrammatic form in Fig.\ \ref{fig1}. 
%


1) We first consider the case ${\hat z}_{\alpha}' \not= 0$ and 
${\hat z}_{\alpha} \not= 0$, or equivalently $q_{\alpha},q_{\alpha}' \not={\tilde q}_{\alpha}$. 
Since the charges of all three particles are assumed to be of equal sign, i.e., all Coulomb potentials are repulsive, 
the three-body Coulomb resolvent has the
spectral representations given in Eqs.\ (I.82) and (I.83) in terms of the scattering wave function $\Psi^{C(+)}_{{\rm {\bf {k}}}_{\alpha}^{0}, 
{\rm {\bf {q}}}_{\alpha}^{0}}({\rm {\bf {P}}})$ for three charged particles in the continuum (${\rm {\bf {P}}} = \{{\rm {\bf {k}}}_{\nu}, 
{\rm {\bf {q}}}_{\nu}\}, \, \nu=1,2,3,$ is the six-dimensional momentum vector). 
Using the latter we can repeat all the steps performed
in Part I, thereby arriving at \cite{note}
\begin{eqnarray}
{\tilde {\cal V}}^{(\beta \beta)}_{\alpha \alpha}
({\rm {\bf{q}}}_{\alpha}', {\rm {\bf{q}}}_{\alpha};z) &=&
4\mu_{\alpha}^{2} \int 
{\frac{d{\rm {\bf {k}}}_{\beta}'''}{({2 \pi})^{3}}} 
\,\int {\frac{d{\rm {\bf {q}}}_{\beta}'''}{({2 \pi})^{3}}} 
\, \int {\frac{d{\rm {\bf {k}}}_{\beta}''}{({2 \pi})^{3}}} 
\,\int {\frac{d{\rm {\bf {q}}}_{\beta}''}{({2 \pi})^{3}}}\,
\int {\frac{d{\rm {\bf {k}}}_{\alpha}^{0}}{({2 \pi})^{3}}} 
\,\int {\frac{d{\rm {\bf {q}}}_{\alpha}^{0}}{({2 \pi})^{3}}}\,
\nonumber \\
&&\times \frac {{\tilde \phi}^{*}_{\alpha}\left(\epsilon_{\alpha \beta } 
({\rm {\bf {q}}}_{\beta}''' + \lambda_{\beta \gamma}
{\rm {\bf {q}}}_{\alpha}')\right)}{\left[({\rm {\bf {q}}}_{\beta}''' + 
\lambda_{\beta \gamma}{\rm {\bf {q}}}_{\alpha}')^{2} - 2\mu_{\alpha}
{\hat z}_{\alpha}' \right]^{1 - i{\hat \eta}_{\alpha}'}}\, \
\frac{4{\pi}e_{\alpha}e_{\gamma}}
{\left[{\rm {\bf {k}}}_{\beta}'''- \epsilon_{\beta \alpha } 
({\rm {\bf {q}}}_{\alpha}' + \lambda_{\alpha \gamma}
{\rm {\bf {q}}}_{\beta}''')\right]^{2}} \nonumber \\
&&\times 
\frac{\Psi^{C(+)}_{{\rm {\bf {k}}}_{\alpha}^{0}, 
{\rm {\bf {q}}}_{\alpha}^{0}}({\rm {\bf {k}}}_{\beta}''', 
{\rm {\bf {q}}}_{\beta}''')
\Psi^{C(+)*}_{{\rm {\bf {k}}}_{\alpha}^{0},
{\rm {\bf {q}}}_{\alpha}^{0}}({\rm {\bf {k}}}_{\beta}'', 
{\rm {\bf {q}}}_{\beta}'')}
{\left[z - {k^{{0}^2}_{\alpha}}/{2\mu_{\alpha}} - {q^{{0}^2}_{\alpha}}/{2M_{\alpha}}\right]} \nonumber \\
&&\times \frac{4{\pi}e_{\alpha}e_{\gamma}}
{\left[{\rm {\bf {k}}}_{\beta}'' - \epsilon_{\beta \alpha}
({\rm {\bf {q}}}_{\alpha} + \lambda_{\alpha \gamma} 
{\rm{\bf{q}}}_{\beta}'')\right]^{2}}
\frac {{\tilde \phi}_{\alpha}\left( \epsilon_{\alpha \beta}( {\rm {\bf 
{q}}}_{\beta}'' + 
\lambda_{\beta \gamma}{\rm {\bf {q}}}_{\alpha})\right)}
{\left[ ({\rm {\bf {q}}}_{\beta}'' + 
\lambda_{\beta \gamma}{\rm {\bf {q}}}_{\alpha})^{2}
- 2\mu_{\alpha} {\hat z}_{\alpha} \right]^{1 - 
i{\hat \eta}_{\alpha}}}.\label{tildint6}
\end{eqnarray}

In contrast to the simpler case considered in Sec.\ \protect\ref{ss4}, the leading singularity of 
${\tilde {\cal V}}^{(\beta \beta)}_{\alpha \alpha}
({\rm {\bf{q}}}_{\alpha}', {\rm {\bf{q}}}_{\alpha};z)$
emerges as the result of the coincidence not only of 
the singularities of the Fourier transforms of the Coulomb potentials 
but also of the forward-scattering 
singularities of the three-body Coulomb scattering wave functions. 
As was shown in Part I (cf.\ Appendix I.D), in the region of integration which 
is relevant for generating the leading singularity, only the leading 
term $\Psi^{C,as(+)'*}_{{\rm {\bf {k}}}_{\alpha}^{0},
{\rm {\bf {q}}}_{\alpha}^{0}}({\rm {\bf {r}}}_{\beta}, 
{\bbox {\rho}}_{\beta})$ in the asymptotic expansion of the three-body coordinate-space 
Coulomb scattering wave function $\Psi^{C(+)*}_{{\rm {\bf {k}}}_{\alpha}^{0},
{\rm {\bf {q}}}_{\alpha}^{0}}({\rm {\bf {r}}}_{\beta}, 
{\bbox {\rho}}_{\beta})$ enters which is known in analytic form
and is given in Eq.\ (I.113). When substituting its momentum space 
representation (I.117) into Eq.\ (\ref{tildint6})
one encounters an expression of the following type: 
\begin{eqnarray}
J_{\alpha}:&=& \, \int {\frac{d{\rm {\bf {k}}}_{\beta}''}{({2 \pi})^{3}}} 
\,\int {\frac{d{\rm {\bf {q}}}_{\beta}''}{({2 \pi})^{3}}}\,
\Psi^{C,as(+)'*}_{{\rm {\bf {k}}}_{\alpha}^{0},
{\rm {\bf {q}}}_{\alpha}^{0}}({\rm {\bf {k}}}_{\beta}'', 
{\rm {\bf {q}}}_{\beta}'')
 \frac{4{\pi}e_{\alpha}e_{\gamma}}
{\left[{\rm {\bf {k}}}_{\beta}'' - \epsilon_{\beta \alpha}
({\rm {\bf {q}}}_{\alpha} + \lambda_{\alpha \gamma} 
{\rm{\bf{q}}}_{\beta}'')\right]^{2}}\nonumber \\
&&\times 
\frac {{\tilde \phi}_{\alpha}\left( \epsilon_{\alpha \beta}( {\rm {\bf 
{q}}}_{\beta}'' + 
\lambda_{\beta \gamma}{\rm {\bf {q}}}_{\alpha})\right)}
{\left[ ({\rm {\bf {q}}}_{\beta}'' + 
\lambda_{\beta \gamma}{\rm {\bf {q}}}_{\alpha})^{2}
- 2\mu_{\alpha} {\hat z}_{\alpha} \right]^{1 - 
i{\hat \eta}_{\alpha}}} \nonumber \\
&=& \, \int {\frac{d{\rm {\bf {q}}}_{\alpha}''}{({2\pi})^{3}}} 
\,\int {\frac{d{\rm {\bf {q}}}_{\beta}''}{({2 \pi})^{3}}}\,
\int {\frac{d{\rm {\bf {k}}}}{({2 \pi})^{3}}} 
\psi^{C(+)*}_{{\rm {\bf {k}}}_{\beta}^{0}}({\rm {\bf k}})
\psi^{C(+)*}_{{\rm {\bf {k}}}^{0}_{\gamma}}({\rm {\bf k}} 
+ {\rm {\bf k}}^{0}_{\gamma} - {\rm {\bf k}}_{\beta}^{0} 
+ \epsilon_{\alpha \beta}({\rm {\bf q}}_{\alpha}''
- {\rm {\bf q}}_{\alpha}^{0})) \nonumber \\
&&\times \psi^{C(+)*}_{{\rm {\bf k}}_{\alpha}^{0}}
({\rm {\bf k}} + {\rm {\bf k}}_{\alpha}^{0} - 
{\rm {\bf k}}^{0}_{\beta} - \epsilon_{\beta \alpha}
({\rm {\bf q}}_{\alpha}'' + {\rm {\bf q}}_{\beta}'' + 
{\rm {\bf q}}_{\gamma}^{0})) 
 \frac{4{\pi}e_{\alpha}e_{\gamma}}
{\left({\rm {\bf {q}}}_{\alpha}'' - {\rm {\bf {q}}}_{\alpha}\right)^{2}} 
\nonumber \\ &&\times
\frac {{\tilde \phi}_{\alpha}\left( \epsilon_{\alpha \beta}( {\rm {\bf 
{q}}}_{\beta}'' + 
\lambda_{\beta \gamma}{\rm {\bf {q}}}_{\alpha})\right)}
{\left[ ({\rm {\bf {q}}}_{\beta}'' + 
\lambda_{\beta \gamma}{\rm {\bf {q}}}_{\alpha})^{2}
- 2 \mu_{\alpha}\,{\hat z}_{\alpha} \right]^{1 - 
i{\hat \eta}_{\alpha}}}, \label{J1}
\end{eqnarray}
where to arrive at the second equality a change of the integration 
variable has been performed. 

We are looking for the behavior of 
$J_{\alpha}$ when the singularities of the integrand at 
\begin{equation}
{\rm {\bf q}}_{\alpha}''- {\rm {\bf q}}_{\alpha}=0 \label{singj1}
\end{equation}
and at
\begin{equation}
({\rm {\bf {q}}}_{\beta}'' + 
\lambda_{\beta \gamma}{\rm {\bf {q}}}_{\alpha})^{2}
- 2 \mu_{\alpha}\,{\hat z}_{\alpha}= 0 \label{singj11}
\end{equation}
collide with the forward-scattering singularities of the wave function 
$\Psi^{C,as(+)'*}_{{\rm {\bf {k}}}_{\alpha}^{0},
{\rm {\bf {q}}}_{\alpha}^{0}}({\rm {\bf {k}}}_{\beta}'', 
{\rm {\bf {q}}}_{\beta}'')$ which occur at 
\begin{equation}
{\rm {\bf q}}_{\nu}''
- {\rm {\bf q}}^{0}_{\nu}=0, \quad \nu=1,\,2,\,3. 
\label{singpsi}
\end{equation}
This will generate the leading singularity of $J_{\alpha}$
in the $(\Delta_{\alpha}^{0})^{2}-$plane where ${\bbox {\Delta}}_{\alpha}^{0}$ has been defined in Eq.\ (\ref{sing04}). And the latter will
eventually give rise to the leading singularity
of ${\tilde {\cal V}}^{(\beta \beta)}_{\alpha \alpha}
({\rm {\bf{q}}}_{\alpha}', {\rm {\bf{q}}}_{\alpha};z)$
in $\Delta_{\alpha}^{2}-$plane. 
 
According Appendix I.E, near the leading singularity $J_{\alpha}$ 
can be written, for $q_{\alpha} \not={\tilde q}_{\alpha}$, as
\begin{eqnarray}
J_{\alpha} &=& 4\pi e_{\alpha}e_{\gamma}
e^{-\pi(\eta_{\alpha}^{0} + \eta_{\beta}^{0} + \eta_{\gamma}^{0})/2}
\frac{\Gamma(1 - i({\hat \eta}_{\alpha} + \eta_{\alpha}^{0}))\,
\Gamma(1 - i(\eta_{\beta}^{0} + \eta_{\gamma}^{0}))}
{\Gamma(1 - i{\hat \eta}_{\alpha})} \nonumber \\
&& \times 
\left[\lambda_{\beta \gamma}^{2}\,(\Delta_{\alpha}^{0})^{2} - \left(k_{\alpha}^{0} - i \sqrt{-2 \mu_{\alpha} {\hat z}_{\alpha}^{*}}\right)^{2}
\right]^{-i\eta_{\alpha}^{0}} 
\left[2\epsilon_{\alpha \beta}
{\mbox {\boldmath $\Delta$}}_{\alpha}^{0} \cdot 
{\rm {\bf k}}_{\beta}^{0}\right]^{-i\eta_{\beta}^{0}} 
{\lbrack - 2\epsilon_{\alpha \beta}\,
{\mbox {\boldmath $\Delta$}}_{\alpha}^{0} \cdot {\rm {\bf k}}_{\gamma}^{0} 
\rbrack^{-i\eta_{\gamma}^{0}}} 
\nonumber \\
&& \times \frac{1}{\lbrack -2 \epsilon_{\alpha \beta}\,
\lambda_{\beta \gamma}\, {\mbox {\boldmath $\Delta$}}_{\alpha}^{0}
\cdot {\rm {\bf k}}_{\alpha}^{0}
+ \sigma_{\alpha}^{0}(q_{\alpha};z)\rbrack^{1 - 
i({\hat \eta}_{\alpha} + \eta_{\alpha}^{0})}}\,
\frac{{\tilde J}_{\alpha}}
{[({\Delta_{\alpha}^{0}})^{2}]^{1 - i(\eta_{\alpha}^{0} + 
\eta_{\gamma}^{0} )}} , 
\label{J1s}
\end{eqnarray}
where ${\tilde J_{\alpha}}$ remains finite at $\Delta_{\alpha}^{0} =
0$. Here, we have introduced the abbreviation 
\begin{equation}
\sigma_{\alpha}^{0}(q_{\alpha};z) =
{k_{\alpha}^{0}}^{2} - 2\,\mu_{\alpha}{\hat z}_{\alpha}.\label{szr1}
\end{equation}
Moreover, terms $\sim O(({\Delta_{\alpha}^{0}})^{2})$ have already been omitted as compared to $O({\Delta_{\alpha}^{0}})$. The only difference between 
Eq.\ (I.E.32) and (\ref{J1s}) concerns the identity 
\begin{equation}
\epsilon_{\alpha \beta}\,{\mbox {\boldmath $\Delta$}}_{\beta}^{0} +
{\rm {\bf k}}_{\alpha} = {\rm {\bf k}}_{\alpha}^{0} - 
\epsilon_{\alpha \beta}\,\lambda_{\beta \gamma}\,
 {\mbox {\boldmath $\Delta$}}_{\alpha}^{0} \label{rl1}
\end{equation}
which follows from the standard relations between the various 
momenta (cf.\ Appendix I.A) and has been used in deriving 
(\ref{J1s}). 
The Coulomb parameters $\eta_{\nu}^{0}, \, \nu=1,2,3,$ are defined in Eq.\ (I.123), and 
$\Gamma(z)$ is the Gamma function. 

Similarly, for $q_{\alpha}' \not= {\tilde q}_{\alpha}$, 
the leading singularity of 
\begin{eqnarray}
J_{\alpha}'^{*}:&=& \, 
\int {\frac{d{\rm {\bf {q}}}_{\alpha}'''}{({2\pi})^{3}}} 
\,\int {\frac{d{\rm {\bf {q}}}_{\beta}'''}{({2 \pi})^{3}}}\,
\int {\frac{d{\rm {\bf {k}}}}{({2 \pi})^{3}}} 
\psi^{C(+)}_{{\rm {\bf {k}}}_{\beta}^{0}}({\rm {\bf k}})
\psi^{C(+)}_{{\rm {\bf {k}}}^{0}_{\gamma}}({\rm {\bf k}} 
+ {\rm {\bf k}}^{0}_{\gamma} - {\rm {\bf k}}_{\beta}^{0} 
+ \epsilon_{\alpha \beta}({\rm {\bf q}}_{\alpha}'''
- {\rm {\bf q}}_{\alpha}^{0})) \nonumber \\
&&\times \psi^{C(+)}_{{\rm {\bf k}}_{\alpha}^{0}}
({\rm {\bf k}} + {\rm {\bf k}}_{\alpha}^{0} - 
{\rm {\bf k}}^{0}_{\beta} - \epsilon_{\beta \alpha}
({\rm {\bf q}}_{\alpha}''' + {\rm {\bf q}}_{\beta}''' + 
{\rm {\bf q}}_{\gamma}^{0})) 
 \frac{4{\pi}e_{\alpha}e_{\gamma}}
{\left({\rm {\bf {q}}}_{\alpha}''' - {\rm {\bf {q}}}_{\alpha}'\right)^{2}} 
\nonumber \\ &&\times
\frac {{\tilde \phi}_{\alpha}^{*}
\left( \epsilon_{\alpha \beta}( {\rm {\bf q}}_{\beta}''' + 
\lambda_{\beta \gamma}{\rm {\bf {q}}}_{\alpha}')\right)}
{\left[ ({\rm {\bf {q}}}_{\beta}''' + 
\lambda_{\beta \gamma}{\rm {\bf q}}_{\alpha}')^{2}
- 2 \mu_{\alpha}\,{\hat z}_{\alpha}' \right]^{1 - 
i{\hat \eta}_{\alpha}'}}, \label{J12}
\end{eqnarray}
is generated by the coincidence
of the zeroes of the denominators which occur at 
\begin{equation}
{\rm {\bf {q}}}_{\alpha}'''- {\rm {\bf {q}}}_{\alpha}'=0 =
({\rm {\bf {q}}}_{\beta}''' + 
\lambda_{\alpha \gamma}{\rm {\bf {q}}}_{\alpha}')^{2} -
2\, \mu_{\alpha}\,{\hat z}_{\alpha}', \label{singj2}
\end{equation}
with the forward-scattering singularities of the wave function
$\Psi^{C,as(+)'}_{{\rm {\bf {k}}}_{\alpha}^{0},
{\rm {\bf {q}}}_{\alpha}^{0}}({\rm {\bf {k}}}_{\beta}''', 
{\rm {\bf {q}}}_{\beta}''')$ at
\begin{equation}
{\rm {\bf q}}_{\nu}'''- {\rm {\bf q}}^{0}_{\nu}=0, \quad \nu=1,\,2,\,3. 
\label{singpsi1}
\end{equation}
In its vicinity leading singular term is of the form
\begin{eqnarray}
J_{\alpha}'^{*} &=& 4\pi e_{\alpha}e_{\gamma}
e^{-\pi(\eta_{\alpha}^{0} + \eta_{\beta}^{0} + \eta_{\gamma}^{0})/2}
\frac{\Gamma(1 - i({\hat \eta}_{\alpha}'- \eta_{\alpha}^{0}))\,
\Gamma(1 + i(\eta_{\beta}^{0} + \eta_{\gamma}^{0}))}
{\Gamma(1 - i{\hat \eta}_{\alpha}')} \nonumber \\
&& \times 
\left[\lambda_{\beta \gamma}^{2}\,(\Delta_{\alpha}'^{0})^{2} - \left(k_{\alpha}^{0} - i \sqrt{-2 \mu_{\alpha} {\hat z}_{\alpha}'^{*}}\right)^{2} \right]^{i\eta_{\alpha}^{0}} 
\left[2\epsilon_{\alpha \beta}
{\mbox {\boldmath $\Delta$}}_{\alpha}'^{0} \cdot 
{\rm {\bf k}}_{\beta}^{0}\right]^{i\eta_{\beta}^{0}} 
{\lbrack - 2\epsilon_{\alpha \beta}\,
{\mbox {\boldmath $\Delta$}}_{\alpha}'^{0} 
\cdot {\rm {\bf k}}_{\gamma}^{0} 
\rbrack^{i\eta_{\gamma}^{0}}} 
\nonumber \\
&& \times \frac{1}{\lbrack -2 \epsilon_{\alpha \beta}\,
\lambda_{\beta \gamma}\, {\mbox {\boldmath $\Delta$}}_{\alpha}'^{0}
\cdot {\rm {\bf k}}_{\alpha}^{0}
+ \sigma_{\alpha}^{0}(q_{\alpha}';z)\rbrack^{1 - 
i({\hat \eta}_{\alpha}' - \eta_{\alpha}^{0})}}\,
\frac{1}{[({\Delta_{\alpha}'^{0}})^{2}]^{1 + i(\eta_{\alpha}^{0} + 
\eta_{\gamma}^{0} )}} {\tilde J}_{\alpha}'^{*}, 
\label{Jp1s}
\end{eqnarray}
where ${\tilde J_{\alpha}'^{*}}$ remains finite 
at ${\mbox {\boldmath $\Delta$}}_{\alpha}'^{0} = 0$. 
Here, $\sigma_{\alpha}^{0}(q_{\alpha}';z)=
{k_{\alpha}^{0}}^{2} - 2\,\mu_{\alpha}\,{\hat z}_{\alpha}'$. 

Taking into account (\ref{J1s}) and (\ref{Jp1s}), and expressing ${\mbox {\boldmath $\Delta$}}_{\alpha}'^{0} $ according to (\ref{sing04}) - (\ref{dltsng1}), we derive 
from (\ref{tildint6}) for the leading singular part in the limit $\Delta_{\alpha} \to 0$ of 
${\tilde {\cal V}}^{(\beta \beta)}_{\alpha \alpha}
({\rm {\bf{q}}}_{\alpha}', {\rm {\bf{q}}}_{\alpha};z)$:
\begin{eqnarray}
{\tilde {\cal V}}^{(\beta \beta)(s)}_{\alpha \alpha}
({\rm {\bf{q}}}_{\alpha}', {\rm {\bf{q}}}_{\alpha};z) &=& 4 \mu_{\alpha}^2
\int {\frac{d{\rm {\bf {k}}}_{\alpha}^{0}}{({2 \pi})^{3}}}
\,\int {\frac{d{\rm {\bf{q}}}_{\alpha}^{0}}{({2 \pi})^{3}}}\,
\frac{J_{\alpha}'^{*}\,J_{\alpha}}
{\left[z - {k^{0}_{\alpha}}^{2}/2\mu_{\alpha} -
{q^{0}_{\alpha}}^{2}/2M_{\alpha}\right]} 
\label{JS21} \\
&=& \frac{64{\pi}^{2}\,e_{\alpha}^{2}\,e_{\gamma}^{2}\,\mu_{\alpha}^2}{\Gamma(1 - i{\hat \eta}_{\alpha}')\,
\Gamma(1 - i{\hat \eta}_{\alpha})} 
\int {\frac{d{\rm {\bf k}}_{\alpha}^{0}}{({2 \pi})^{3}}} \,
\int {\frac{d{\mbox {\boldmath $\Delta$}}_{\alpha}^{0}}{({2 \pi})^{3}}}
e^{-\pi(\eta_{\alpha}^{0} + \eta_{\beta}^{0} + 
\eta_{\gamma}^{0})} \nonumber \\
&& \times \Gamma(1 - i({\hat \eta}_{\alpha}' - 
\eta_{\alpha}^{0}))\, \Gamma(1 - i({\hat \eta}_{\alpha} + \eta_{\alpha}^{0}))\,
\mid \Gamma(1 - i(\eta_{\beta}^{0} + \eta_{\gamma}^{0}))\mid^{2}\, \nonumber \\
&& \times 
\left[\frac{\lambda_{\beta \gamma}^{2}\,({\mbox {\boldmath $\Delta$}}_{\alpha}^{0} - 
{\mbox {\boldmath $\Delta$}}_{\alpha})^{2} - \left(k_{\alpha}^{0} - i \sqrt{-2 \mu_{\alpha} {\hat z}_{\alpha}'^{*}}\right)^{2} }{\lambda_{\beta \gamma}^{2}\,(\Delta_{\alpha}^{0})^{2} - \left(k_{\alpha}^{0} - i \sqrt{-2 \mu_{\alpha} {\hat z}_{\alpha}^{*}}\right)^{2} }\right]^{i\eta_{\alpha}^{0}} \nonumber \\
&&\times 
\left[ 
{1 - {\mbox {\boldmath $\Delta$}}_{\alpha} \cdot 
{\rm {\bf k}}_{\beta}^{0}} / {({\mbox {\boldmath $\Delta$}}_{\alpha}^{0} \cdot 
{\rm {\bf k}}_{\beta}^{0})}\right]^{i \eta_{\beta}^{0}} 
\left[ 1 - {
{\mbox {\boldmath $\Delta$}}_{\alpha} 
\cdot {\rm {\bf k}}_{\gamma}^{0}}/ {({\mbox {\boldmath $\Delta$}}_{\alpha}^{0} 
\cdot {\rm {\bf k}}_{\gamma}^{0})} \right]^{i \eta_{\gamma}^{0}} 
\nonumber \\
&& \times \frac{1}{\lbrack -2 \epsilon_{\alpha \beta}\,
\lambda_{\beta \gamma}\, ({\mbox {\boldmath $\Delta$}}_{\alpha}^{0} - 
{\mbox {\boldmath $\Delta$}}_{\alpha})
\cdot {\rm {\bf k}}_{\alpha}^{0}
+ \sigma_{\alpha}^{0}(q_{\alpha}';z)\rbrack^{1 -
i({\hat \eta}_{\alpha}' - \eta_{\alpha}^{0})}}\,\nonumber \\
&&\times 
\frac{1}{[ ({\mbox {\boldmath $\Delta$}}_{\alpha}^{0} - 
{\mbox {\boldmath $\Delta$}}_{\alpha})^{2}]^{1 + i(\eta_{\alpha}^{0} + 
\eta_{\gamma}^{0} )}} \nonumber \\
&&\times 
\frac{{\tilde J}_{\alpha}'^{*}\,{\tilde J}_{\alpha}}
{\left[{\hat z}_{\alpha} - 
{\mbox {\boldmath $\Delta$}}_{\alpha}^{0} \cdot
\left({\mbox {\boldmath $\Delta$}}_{\alpha}^{0} + 2\,{\rm {\bf q}}_{\alpha} \right)/
2\,M_{\alpha} - {k_{\alpha}^{0}}^{2}/2\,\mu_{\alpha}\right]}
\nonumber \\
&& \times \frac{1}{\lbrack -2 \epsilon_{\alpha \beta}\,
\lambda_{\beta \gamma}\, {\mbox {\boldmath $\Delta$}}_{\alpha}^{0}
\cdot {\rm {\bf k}}_{\alpha}^{0}
+ \sigma_{\alpha}^{0}(q_{\alpha};z)\rbrack^{1 - 
i({\hat \eta}_{\alpha} + \eta_{\alpha}^{0})}}\,
\frac{1}{[({\Delta_{\alpha}^{0}})^{2}]^{1 - i(\eta_{\alpha}^{0} + 
\eta_{\gamma}^{0} )}}. 
\label{Jp2s1}
\end{eqnarray}

It is instructive to compare the integrals over
${\mbox {\boldmath $\Delta$}}_{\alpha}^{0}$ in 
Eqs.\ (\ref{tildint3}) and (\ref{Jp2s1}). 
Inspection reveals that differences arise from the presence 
of extra factors $\sim \Delta_{\alpha}/\Delta_{\alpha}^{0}$ 
, and from the Coulomb distortion 
of the poles in Eq.\ (\ref{tildint3}) 
at ${\mbox {\boldmath $\Delta$}}_{\alpha}^{0}=0$
and ${\mbox {\boldmath $\Delta$}}_{\alpha}^{0}-
{\mbox {\boldmath $\Delta$}}_{\alpha}=0$ 
into branch points. 
This latter effect is due to the presence in 
${\tilde {\cal V}}^{(\beta \beta)}_{\alpha \alpha}
({\rm {\bf{q}}}_{\alpha}', {\rm {\bf{q}}}_{\alpha};z)$
of the three-body Coulomb Green's function describing the 
propagation of the three charged particles $\alpha, \, \beta,$ 
and $\gamma,$ in the intermediate state. 
Now, making the scaling transformation
(\ref{sctd1}) we immediately arrive at the final result:
\begin{equation}
{\tilde {\cal V}}^{(\beta \beta)(s)}_{\alpha \alpha}
({\rm {\bf{q}}}_{\alpha}', {\rm {\bf{q}}}_{\alpha};z) 
\stackrel{\Delta_{\alpha} \to 0}{=} 
\frac{1}{\Delta_{\alpha}}\,J_{\alpha \alpha}^{(\beta \beta)} , \label{vsaa1}
\end{equation}
whith
\begin{eqnarray}
&&J_{\alpha \alpha}^{(\beta \beta)} = -
\frac{128{\pi}^{2}\mu_{\alpha}^{3}\,e_{\alpha}^{2}e_{\gamma}^{2}}
{\Gamma^{2}(1 - i{\hat \eta}_{\alpha})}\, 
\int {\frac{d{\rm {\bf k}}_{\alpha}^{0}}{({2 \pi})^{3}}} \,
\int {\frac{d{\rm {\bf v}}_{\alpha}}{({2 \pi})^{3}}}
\; e^{-\pi(\eta_{\alpha}^{0} + \eta_{\beta}^{0} + 
\eta_{\gamma}^{0})} \nonumber \\
&& \times \Gamma \left(1 - i({\hat \eta}_{\alpha} - 
\eta_{\alpha}^{0})\right)\, \Gamma \left(1 - i({\hat \eta}_{\alpha} + \eta_{\alpha}^{0})\right)\,
\left|\Gamma \left(1 - i(\eta_{\beta}^{0} + \eta_{\gamma}^{0})\right) \right|^{2}\, 
\nonumber \\
&& \times 
\left[ 1-{{\hat {\bbox{\Delta}}}_{\alpha} \cdot 
{\rm {\bf k}}_{\beta}^{0}} /
{({\rm {\bf v}}_{\alpha} \cdot 
{\rm {\bf k}}_{\beta}^{0})}\right]^{i \eta_{\beta}^{0}} 
\left[ 1- 
{{\hat {\bbox{\Delta}}}_{\alpha}\cdot {\rm {\bf k}}_{\gamma}^{0}} /
{({\rm {\bf v}}_{\alpha} 
\cdot {\rm {\bf k}}_{\gamma}^{0})} \right]^{i \eta_{\gamma}^{0}} 
\nonumber \\
&& \times 
\frac{1}{({\rm {\bf v}}_{\alpha} - 
{\hat {\mbox {\boldmath $\Delta$}}}_{\alpha})^{2 + 2i(\eta_{\alpha}^{0} + 
\eta_{\gamma}^{0} )}}\; \frac{1}{v_{\alpha}^{2 - 2i(\eta_{\alpha}^{0} + 
\eta_{\gamma}^{0} )}} \;
\frac{{\tilde J}_{\alpha}'^{*}\,{\tilde J}_{\alpha}}
{\left[{k_{\alpha}^{0}}^{2} - 2\,\mu_{\alpha}\,{\hat z}_{\alpha} \right]^{3 - 2i {\hat \eta}_{\alpha}}}.
\label{Jp2s2}
\end{eqnarray}
Here, we have taken into account that $\lim_{\Delta_{\alpha} \to 0}{\hat z}_{\alpha}' = {\hat z}_{\alpha} $.

Let us comment on this result. 
In our previous work 
on two-charged particle scattering \cite{am94}, 
the influence of the two-body Coulomb Green's function
on the analytic behavior of the two-particle Coulomb 
scattering amplitude defined via $T^C = V^C + V^C G^C V^C$ 
had been investigated, by using the spectral representation of $G^C$. 
There we had come up with the following fundamental result:
due to the mutual cancellation of the Coulomb distortion factors 
in the pair of complex conjugate Coulomb scattering wave functions, 
the presence of the two-body Coulomb Green's
function in the intermediate state did not affect the leading
singular behavior of the Coulomb scattering amplitude
at small momentum transfer as compared to the lowest-order 
contribution in a Born series expansion of the latter (i.e., 
if $G^C$ is replaced by $ G_0$ in the second term on the 
r.\ h.\ side). 

Now we have encountered
the analogous situation in the three-body case 
which, of course, is more complicated. When looking
for the leading singular term of 
${\tilde {\cal V}}^{(\beta \beta)}_{\alpha \alpha}
({\rm {\bf{q}}}_{\alpha}', {\rm {\bf{q}}}_{\alpha};z)$ (see Eq.\ (\ref{tildint6})), in the spectral representation 
of the Coulomb Green's function, 
the exact three-body Coulomb wave 
functions could be approximated by their leading asymptotic 
parts. Since the (asymptotic) Coulomb wave functions appear there 
as a pair of complex conjugates of each other, the phase factors
in $J_{\alpha}$ and $J_{\alpha}'^{*}$ containing the Coulomb parameters $\eta_{\nu}^{0},\, \nu= \alpha,\,\beta,\, \gamma$, 
are complex conjugates as well. This fact eventually led to 
the same result: 
the presence of the three-body Coulomb Green's function in the 
intermediate state of the exact effective potential part 
${\tilde {\cal V}}_{\alpha \alpha}^{(\beta \beta)}
({\rm {\bf{q}}}_{\alpha}', {\rm {\bf{q}}}_{\alpha};z) $, Eq.\ (\ref{tildfp2}), does not affect the singular behavior at small momentum transfer, as compared to its lowest-order contribution ${\tilde {\cal V}}_{\alpha \alpha}^{(2)(\beta \beta)}
({\rm {\bf{q}}}_{\alpha}', {\rm {\bf{q}}}_{\alpha};z)$ given in Eq.\ (\ref{tildfp20}).

2) Next assume $q_{\alpha} = {\tilde q}_{\alpha}$. 
The leading singular term of $J_{\alpha}$ is
given by Eq.\ (I.E.33), 
and analogously that of $J_{\alpha}'$ by 
\begin{eqnarray}
J_{\alpha}'^{*}\, &=& \, 4\pi e_{\alpha}e_{\gamma}\, 
e^{-\pi(\eta_{\alpha}^{0} + \eta_{\beta}^{0} 
+ \eta_{\gamma}^{0})/2}\,
\Gamma(1 + i \eta_{\alpha}^{0})\,\Gamma\left(1 +
i(\eta_{\beta}^{0} + \eta_{\gamma}^{0})\right) \nonumber \\
&& \times \left[2\epsilon_{\alpha \beta}
{\mbox {\boldmath $\Delta$}}_{\alpha}'^{0} \cdot 
{\rm {\bf k}}_{\beta}^{0}\right]^{i\eta_{\beta}^{0}} 
{\lbrack - 2\epsilon_{\alpha \beta}\,
{\mbox {\boldmath $\Delta$}}_{\alpha}'^{0} \cdot 
{\rm {\bf k}}_{\gamma}^{0} \rbrack^{i\eta_{\gamma}^{0}}} 
\frac{{\tilde L}_{1}'^{*}}{\lbrack ({\Delta_{\alpha}'^{0}})^{2} 
\rbrack^{1 + i(\eta_{\beta}^{0} + \eta_{\gamma}^{0})}}. 
\label{jpcc1}
\end{eqnarray}
Correspondingly, taking into account that for $\Delta_{\alpha} = 0$
and $q_{\alpha}={\tilde q}_{\alpha}$ one has
${\hat z}_{\alpha}'= {\hat z}_{\alpha}=0$, 
the leading singular term of 
${\tilde {\cal V}}^{(\beta \beta)}_{\alpha \alpha}
({\rm {\bf{q}}}_{\alpha}', {\rm {\bf{q}}}_{\alpha};z)$ is found to be
\begin{equation}
{\tilde {\cal V}}^{(\beta \beta)(s)}_{\alpha \alpha}
({\tilde {\rm {\bf{q}}}}_{\alpha}', {\tilde {\rm {\bf{q}}}}_{\alpha};z) 
\stackrel{\Delta_{\alpha} \to 0}{=} 
\frac{1}{\Delta_{\alpha}}\,J_{\alpha \alpha}^{(\beta \beta)} , \label{vsaa2}
\end{equation}
with
\begin{eqnarray}
&&J_{\alpha \alpha}^{(\beta \beta)} = 
-128{\pi}^{2}\,\mu_{\alpha}^{3}\,e_{\alpha}^{2}\,e_{\gamma}^{2}\, 
\int {\frac{d{\rm {\bf k}}_{\alpha}^{0}}{({2 \pi})^{3}}}\,
\int {\frac{d{\rm {\bf v}}_{\alpha}^{0}}{({2 \pi})^{3}}}
\nonumber \\
&& \times e^{-\pi(\eta_{\alpha}^{0} + \eta_{\beta}^{0} + 
\eta_{\gamma}^{0})} \left| \Gamma \left(1 - i\eta_{\alpha}^{0}\right) \right|^{2} \, 
\left| \Gamma \left(1 - i(\eta_{\beta}^{0} + \eta_{\gamma}^{0})\right) \right|^{2}\, 
\nonumber \\
&& \times 
\left[ 1-{{\hat {\bbox{\Delta}}}_{\alpha} \cdot 
{\rm {\bf k}}_{\beta}^{0}} /
{({\rm {\bf v}}_{\alpha} \cdot 
{\rm {\bf k}}_{\beta}^{0})}\right]^{i \eta_{\beta}^{0}} 
\left[ 1- 
{{\hat {\bbox{\Delta}}}_{\alpha}\cdot {\rm {\bf k}}_{\gamma}^{0}} /
{({\rm {\bf v}}_{\alpha} 
\cdot {\rm {\bf k}}_{\gamma}^{0})} \right]^{i \eta_{\gamma}^{0}} 
\nonumber \\
&& \times 
\frac{1}{({\rm {\bf v}}_{\alpha} -
{\hat {\mbox {\boldmath $\Delta$}}}_{\alpha})^{2 + 
2 i(\eta_{\alpha}^{0} + \eta_{\gamma}^{0} )}} \;
\frac{1}{v_{\alpha}^{2 - 2 i(\eta_{\alpha}^{0} + 
\eta_{\gamma}^{0} )}} \; \frac{{\tilde L}_{1}'^{*}\,{\tilde L}_{1}}
{{k_{\alpha}^{0}}^{2}}
. 
\label{Jp2s3}
\end{eqnarray}

3) It is not difficult to see that all other contributions ${\tilde {\cal V}}_{\alpha \alpha}^{(\nu \sigma)}
({\rm {\bf{q}}}_{\alpha}', {\rm {\bf{q}}}_{\alpha};z)$, with $\nu,\sigma = \beta,\gamma,$ to the effective potential Eq.\ (\ref{tildfp1}) behave in the limit $\Delta_{\alpha} \to 0$ as described above. Consequently, the leading singular behavior of the full effective potential part (\ref{defpot13}) is given as 
\begin{equation}
{\tilde {\cal V}}_{\alpha \alpha}
({\rm {\bf{q}}}_{\alpha}', {\rm {\bf{q}}}_{\alpha};z) 
\stackrel{\Delta_{\alpha} \to 0}{\sim} 
\frac{1}{\Delta_{\alpha}}.
\end{equation}
This concludes the proof of the Theorem.

\newpage
\section{Concluding Remarks} \protect\label{concl}

The results of the present investigation, together with those of paper I (Phys.\ Rev.\ C {\bf 61}, 064006 (2000)), provide the proof that momentum space three-body integral equations {\em in the form of effective two-body AGS equations} can be used with confidence to calculate all possible arrangement (i.e., $2 \to 2$) amplitudes below and above the three-body threshold, provided the charges of all three particles are of the same sign. However, as information concerning the singularity properties of the effective potentials occurring in the analogous equations for breakup ($2 \to 3$) or even $3 \to 3$ amplitudes is still lacking, the results obtained so far do not yet constitute a proof of compactness of the kernels of the {\em genuine three-body} integral equations of the Faddeev \cite{fad61} or AGS \cite{ags67} type. One obvious consequence is that application of methods which aim at directly solving these latter equations would (as yet) be without mathematical justification.
While reactions of the $3 \to 3$ type are less of practical interest, experimental study of breakup processes is vigorously pursued in many laboratories. Hence, it is of great importance to continue these investigations, at least for the effective potentials occurring in the (integral) equations for $2 \to 3$ amplitudes. Note that, if only two of the three particles are charged, with charges of the same sign, the proofs provided by Alt, Sandhas, and Ziegelmann \cite{asz78,as96} within the screening and renormalisation approach constitute a proof of compactness of the kernels of the corresponding {\em three-body} integral equations (cf.\ the validation of this assertion in \cite{fm85}).

A further comment concerns the practical applicability of the momentum space approach. Indeed, evaluation of the exact effective potential ${\cal V}_{\beta \alpha}({\rm {\bf{q}}}_{\beta}',
{\rm {\bf{q}}}_{\alpha};z) = \langle {\rm {\bf q}}_{\beta}',
\chi_{\beta} \mid G^{C}(z) - \delta_{\beta \alpha}G_{\alpha}^{C}(z) \mid
\chi_{\alpha},{\rm {\bf q}}_{\alpha} \rangle$, Eq.\ (I.11a), appears to be beyond present means, due to the presence of the three-body Coulomb resolvent $G^{C}$. On the other hand, a perturbative calculation of ${\cal V}_{\beta \alpha}({\rm {\bf{q}}}_{\beta}',{\rm {\bf{q}}}_{\alpha};z)$ based on the Neumann series expansion of $G^{C}$ may also be unsatisfactory. For, as Theorem 2 (cf.\ Part I) states for its nondiagonal and the Auxiliary Theorem of the present paper for its diagonal part, even the lowest-order contribution ${\cal V}_{\beta \alpha}^{(2)}({\rm {\bf{q}}}_{\beta}',
{\rm {\bf{q}}}_{\alpha};z) $ and, thus, also all the higher-order terms of the perturbation series have the same leading singularity as the exact expression and consequently contribute to the strength of the residue at the singularity. Therefore, in principle no term of this infinite series should be omitted unless, of course, their contribution to the residue strength turns out to become rapidly smaller as the order of iteration increases.

\end{document}